\documentclass{aa}
\usepackage{graphicx}
\usepackage{txfonts}

\usepackage{color}
\usepackage{array}
\usepackage{caption}
\usepackage{subcaption}
\usepackage{amsmath,amssymb,amsfonts}
\usepackage{gensymb, footnote}
\usepackage{sidecap}
\usepackage{changepage}
\usepackage{tabu}
\usepackage[toc,page]{appendix}
\usepackage{multirow}
\usepackage[]{hyperref}
\hypersetup{
   colorlinks = true,
   citecolor ={cyan}
}

\usepackage{tablefootnote}
\usepackage{threeparttable}
\usepackage{upgreek}

\title{BHs in the Milky Way with Gaia}
\author{Soetkin Janssens}
\date{March 2021}

\newcommand{\Teff}{T_{\text{eff}}}

\newcommand{\Modot}{M_{\odot}}
\newcommand{\MSun}{M_{\odot}}
\newcommand{\Rodot}{R_{\odot}}

\newcommand{\Tabref}{Table\,\ref}
\newcommand{\Figref}{Fig.\,\ref}
\newcommand{\Secref}{Sect.\,\ref}
\newcommand{\Eqref}{Eq.\,\eqref}

\newcommand{\nextline}{\\\indent}

\newcommand{\Gaia}{\textit{Gaia}}
\newcommand{\Gaias}{\textit{Gaia}\space}

\defcitealias{Janssens_2022}{Paper I}
\defcitealias{Shahaf_2019}{SMF19}

\begin{document}
   \title{Detection of single-degenerate massive binaries with Gaia:
   The impact of blue supergiants, triples, mass precision, and high-precision parallax requirements} \titlerunning{Single-degenerate massive binaries with Gaia:
   The impact of several biases}

   \author{S. Janssens \inst{1} 
           \and T. Shenar\inst{2}
           \and H. Sana\inst{1}
           \and P. Marchant\inst{1}
          }

   \institute{Institute of Astronomy, KU Leuven, Celestijnenlaan 200D, 3001 Leuven, Belgium\\ \email{soetkin.janssens@kuleuven.be}
    \and
    Anton Pannekoek Institute for Astronomy, Science Park 904, 1098 XH, Amsterdam, The Netherlands    
          }    

   \date{}

 
  \abstract
   {X-ray quiet single-degenerate massive binaries are notoriously difficult to detect, and only few have been identified so far. Yet, recent investigations have shown that hundreds of black holes with massive main-sequence companions (OB+BHs) might be identifiable from \Gaias astrometry by using the Astrometric Mass-Ratio Function (AMRF).
   }
   {We aim to investigate a number of biases that can have an impact on the astrometric identification of OB+BH binaries, namely the presence of blue supergiant companions (BSG) instead of dwarfs and the presence of additional companions in the system that are not resolved by \Gaia. We also explore the accuracy with which the primary mass needs to be constrained. Moreover, we assess the impact of high-precision constraints on the detection of binaries by using the conservative constraints imposed to release astrometric orbits in the latest \Gaias data release DR3. We then investigate how much less stringent these constraints need to be in order to obtain information on the BH-formation scenario.}
   {We establish a mass-magnitude relation of BSGs and compute from this BSG AMRF curves. A mock population of OB/BSG+BH binaries, non-degenerate binaries with an OB or BSG primary (OB/BSG+MS), and triples with OB and BSG primaries is used to determine the fraction of false-positive identifications and the effect of the BSG mass-magnitude relation. We compare the number of sources with astrometric DR3 orbits in the second Alma Luminous Star catalogue with new predictions on the detection of OB+BHs using the conservative selection criterion used for publishing astrometric orbits in DR3.
   }
   {We show that the addition of systems with BSG primaries is not impacting the fraction of false-positive identifications significantly. Only for triple systems where the outer star is more luminous and a BSG, the usage of the previously established MS curves might result in a high amount of false-positive identifications. However, such systems are expected to be rare. We also demonstrate that the mass of the primary does not need to be accurately known to benefit from the high identification fraction of OB/BSG+BHs, while keeping the fraction of false-positives low. We find that 11 sources have an astrometric binary orbit available in DR3. None of these sources are OB/BSG+BH candidates. This is in line with the new predictions using the \Gaias DR3 selection criterion.}
   {If the evolutionary stage of the primary stars are unknown, the usage of the BSG curves is recommended over the MS curves to avoid high contamination from BSG+MS systems or triples with a more luminous outer star. This way, the false-positive fractions are decreased by an order of magnitude, reaching values $\ll 1\%$. However, the fraction of identifiable OB+BHs is also significantly reduced (from 68\% to 29\%). In case the mass of the primary star is not known, it is possible to use a fixed estimate. The non-detection of astrometric OB/BSG+BH systems cannot be attributed to the underlying BH-formation scenario, but rather to the stringent selection criterion imposed on the parallax relative uncertainty in DR3. A relaxation of this condition would be needed if we are to find the bulk of the OB+BH population with \Gaia. If possible, we propose that the constraint on the relative parallax precision in DR4 should be improved to $n \times (\varpi/\sigma_{\varpi})_{\text{DR3,single}} > n\times 1000/P_{\text{day}}$, with $(\varpi/\sigma_{\varpi})_{\text{DR3,single}}$ the relative parallax precision for the single source solution in DR3 and $n$ the DR4 improvement of the relative precision compared to the DR3 single star solutions. In DR3, this means a 95\% decrease of the original criterion.}


   \maketitle

\section{Introduction}
Double massive main-sequence binaries (OB+OB) are expected to play an important role in the formation of compact object mergers of neutron stars (NS) and black holes (BH) \citep[e.g.][]{Mandel_2022}. A particularly important intermediate evolutionary stage before a merger is a single degenerate phase (either OB+NS or OB+BH), which might be identified through the emission of X-rays \citep[see][for identified BHs]{Walter_2015,Corral-Santana_2016}, called a high-mass X-ray binary (HMXB). This occurs when the OB star has evolved, either almost filling its Roche lobe leading to a wind-fed X-ray binary or filling its Roche-lobe after which copious mass is being accreted onto the BH. This phase in the evolution is however very short-lived compared to the OB+BH phase, which is determined by the main-sequence lifetime of the OB star \citep[][]{Sen_2021,Hiria_2021}. On top of that, OB+BH systems will not always evolve into HMXBs. For example, an OB+BH with $P>10$\,d would never become a HMXB.\nextline
The search for X-ray quiescent OB+BH systems is proving to be highly challenging. In the past two years, reported BHs in spectroscopic studies are challenged by follow-up studies, often leading to explanations that do not involve BHs (to name a few, see for example some discussions on LB-1: \citealt{Liu_2019}, \citealt{Abdul-Masih_2020}, \citealt{El_badry_2020_LB1}, \citealt{Shenar_2020}, \citealt{Irrgang_2020}, \citealt{Liu_2020}, \citealt{Lennon_2021}; on HR6819: \citealt{Bodensteiner_2020}, \citealt{Mazeh_2020}, \citealt{Rivinius_2020}, \citealt{Romagnolo_2021}, \citealt{Frost_2022}; on NGC 1850: \citealt{El_badry_2022}, \citealt{Saracino_2022}, \citealt{Stevance_2022}). Nonetheless, very recent spectroscopic investigations have revealed a handful of such OB+BH systems \citep[e.g.][]{Mahy_2022_BHs,Shenar_2022_Nature}. However, their numbers remain small since these dedicated spectroscopic studies require long time series and high signal-to-noise spectra. \nextline
The \Gaias space mission \citep[][]{Gaia_colab_2016_mission} has great potential for the detection of single-degenerate binaries \citep[e.g.][]{Breivik_2017,Breivik_2019,Mashian_Loeb_2017, Yalinewich_2018, Yamaguchi_2018, Andrews_2019,Wiktorowicz_2019,Shikauchi_2020, Shikauchi_2022}. A method to astrometrically identify such single-degenerate binaries was developed by \citeauthor{Shahaf_2019} (\citeyear{Shahaf_2019}, hereafter referred to as \citetalias{Shahaf_2019}).\nextline
Using the results of detailed binary evolution simulations of \citet{Langer_2020}, \citeauthor{Janssens_2022} (\citeyear{Janssens_2022}, hereafter referred to as \citetalias{Janssens_2022}) showed that the identification method presented in \citetalias{Shahaf_2019} could be adapted to identify a predicted 100-200 OB+BH systems among the sources in the second Alma Luminous Star catalogue \citep[ALS\,II;][]{Gonzalez_2021}. The authors further show that the distributions of the parameters of detected OB+BHs is expected to retain information on the BH-formation physics (the kick and the fallback mass).\nextline
In \citetalias{Janssens_2022}, the authors only considered the impact of unevolved non-degenerate OB+MS binaries on the number of false-positive detections -- systems that do not contain a BH that are misidentified as BHs. However, many OB stars are found in triples \citep{Sana_2014,Moe_2017}, which may also contribute to the amount of false-positives. On top of that, the ALS\,II catalogue is magnitude limited and hence evolved massive stars such as supergiants are also abundant. Supergiants are much brighter than main-sequence dwarfs and their mass-magnitude relations are very different, impacting the used OB+BH-identification method. Hence, supergiants too are a potential cause of many false-positives that has not been quantified so far. Finally, in \citetalias{Janssens_2022}, the authors used the nominal \Gaias instrument precision, while much more stringent criteria were implemented when constructing the recently published \Gaias DR3 catalogue.\nextline 
In this work, we investigate the impact of a different mass-magnitude relation, coming from supergiant primaries, and additional companions on the fractions of identifiable OB+BH systems and false-positives. We also explore how accurate the primary masses need to be known to contain the amount of false-positives. Moreover, we investigate the effect of using the adopted \Gaias selection criteria to publish astrometric orbits in DR3 and compare to the results predicted in \citetalias{Janssens_2022}.\nextline
Section \ref{sec_overview} gives an overview of the method used by \citetalias{Janssens_2022} for the identification of OB+BH systems, together with a summary of the possible biases and uncertainties. The impact of the adopted mass-magnitude relation on the identification fractions, higher-order multiple systems, and the accuracy of the primary mass measurements are discussed in Secs. \ref{section_evolved_primaries} to \ref{sec_mass_measurements}. The results of OB+BH detections with DR3 in comparison to new predictions using the detection criteria made in DR3 are discussed in \Secref{sec_gaia_DR3_results}. We end with a summary in \Secref{sec_summary}.

\begin{figure}
    \centering
    \includegraphics[width = \linewidth]{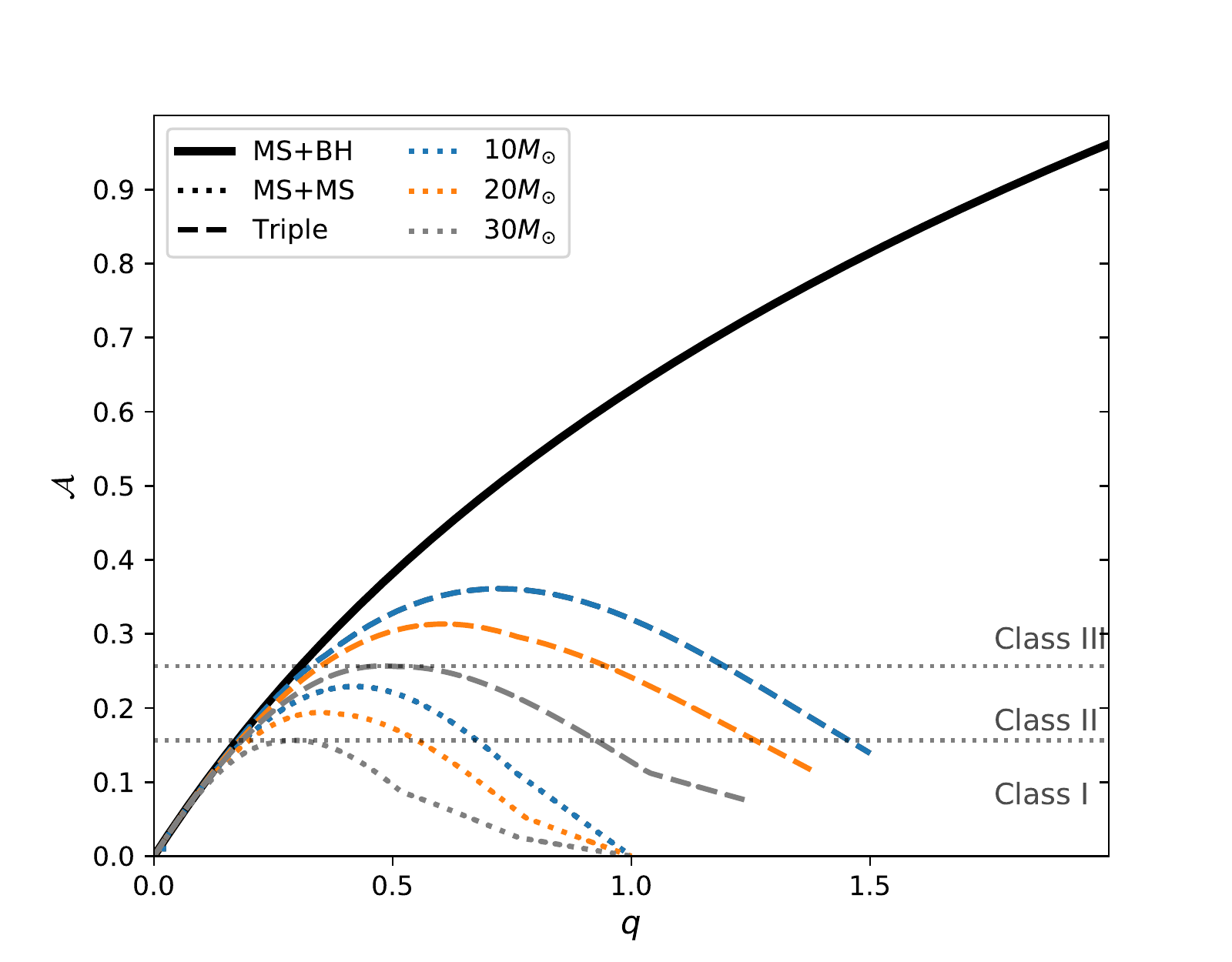}
    \caption{Different theoretical AMRFs for three different primary masses: 10 $\Modot$ (blue), 20 $\Modot$ (orange), and 30 $\Modot$ (grey). The solid black line shows the theoretical AMRF for MS+BH systems, the dotted curves are for OB+MS binaries, and dashed curves are for triple main-sequence dwarf systems. Different classes are also indicated for the 30 $\Modot$ primary by the horizontal dotted lines at the maxima of the triple-AMRF and the binary AMRF (Class III: identifiable OB+BH systems, Class II: triple main-sequence or OB+BH systems, Class I: OB+MS binaries, triple main-sequence systems, or OB+BH binaries).}
    \label{fig_AMRF_curves_SJ22}
\end{figure}

\section{The identification method, possible biases, and uncertainties}\label{sec_overview}
\subsection{The Astrometric Mass-Ratio Function}
The Astrometric Mass-Ratio Function (AMRF; \citetalias{Shahaf_2019}) was adapted to massive stars in \citetalias{Janssens_2022}. The AMRF is a unitless parameter defined as
\begin{equation}\label{eq_AMRF}
    \mathcal{A} = \frac{\alpha}{\varpi} \left( \frac{M_1}{\Modot} \right)^{-1/3} \left( \frac{P}{\text{yr}} \right)^{-2/3} = \frac{q}{(1+q)^{2/3}} \left( 1-\frac{S(1+q)}{q(1+S)} \right),
\end{equation}
where, $\alpha$ is the observed astrometric signal (the observed semi-major axis traced by the photocentre), $\varpi$ is the parallax of the system, $M_1$ is the mass of the primary or most luminous star, and $P$ is the orbital period of the system in years. In the right-most term, $q = M_2/M_1$ is the current mass ratio of the secondary (least luminous) to the primary star and $S=I_2/I_1$ their intensity ratio in the pass band of the instrument measuring the astrometry, in this case the \Gaias G band. From \Eqref{eq_AMRF}, it is clear that the AMRF is sensitive to the adopted mass-luminosity relation through $S$. The right-most term in \Eqref{eq_AMRF} will be referred to as the theoretical AMRF and the middle term as the observational AMRF or $\mathcal{A}$.\nextline
Both \citetalias{Shahaf_2019} and \citetalias{Janssens_2022} consider three different scenarios for the AMRF method: single-degenerate binaries (MS+BH/NS), main-sequence binaries (MS+MS), and triple systems where the outer star is more luminous than the inner binary (triple) since these systems result in the highest AMRF curves (see \Secref{sec_kind_of_triples}). Hence, the inner binary in the triple systems is considered the secondary. Since we are here focusing on massive primaries, the abbreviations above change to OB+BH, OB+MS.\nextline
The theoretical AMRF curves for systems with massive main-sequence dwarf primaries are determined in \citetalias{Janssens_2022} for the three above-mentioned kinds of systems (see \Figref{fig_AMRF_curves_SJ22}) as a function of $q$. However, without a mass estimate for the luminous companion, the mass ratio of the system is unconstrained. We therefore work under the assumption that $q$ is not a known quantity.\nextline
\citetalias{Shahaf_2019} introduced three classes, based on the principle that, without measurement uncertainties, the observational AMRF of any kind of system cannot exceed the theoretical curve for that kind of system (e.g. OB+MS binaries cannot exceed the theoretical curve for OB+MS binaries). Class III systems are systems that are found with an observational AMRF above the maximum of the triple curve. These are hence classified as OB+BH systems. Systems found with an observational AMRF below the maximum of the triple curve but above the maximum of the binary curve are classified as Class II. Here, we find a mix of triples and OB+BH systems. Class I systems are systems with an observational AMRF below the maximum of the binary curve. Class I contains OB+MS systems, but also triples and OB+BH systems. The three classes are marked in \Figref{fig_AMRF_curves_SJ22} only for a 30$\Modot$ primary.

\subsection{Biases previously considered in \citetalias{Janssens_2022}}
The identification criterion for Class III imposed in \citetalias{Janssens_2022} is such that a system requires an observational AMRF above the maximum of the triple curve within one sigma or $\mathcal{A}-\sigma_{\mathcal{A}} > \mathcal{A}_{\text{triple}}$. Although the three classes are in principle well defined, measurement uncertainties might cause observational data points in the diagram to shift into a different class, which impact the identification of single-degenerate binaries through the AMRF method. A proper determination of the true numbers of the expected OB+BHs is important, as it can contain information on the survivability of single-degenerate systems through the collapse of their non-degenerate progenitors (\citetalias{Janssens_2022}).\nextline
On the one hand, OB+MS or triple systems might get falsely identified as OB+BHs. These false identifications are called false-positives. On the other hand, measurement uncertainties can also cause some OB+BH systems to appear as Class II or I systems. Both false-positives and non-detections can affect the number of identifications. As shown in \citetalias{Janssens_2022}, the expected contribution of OB+MS binaries to the amount of false-positives is negligible.

\begin{figure}
    \centering
    \includegraphics[width = \linewidth]{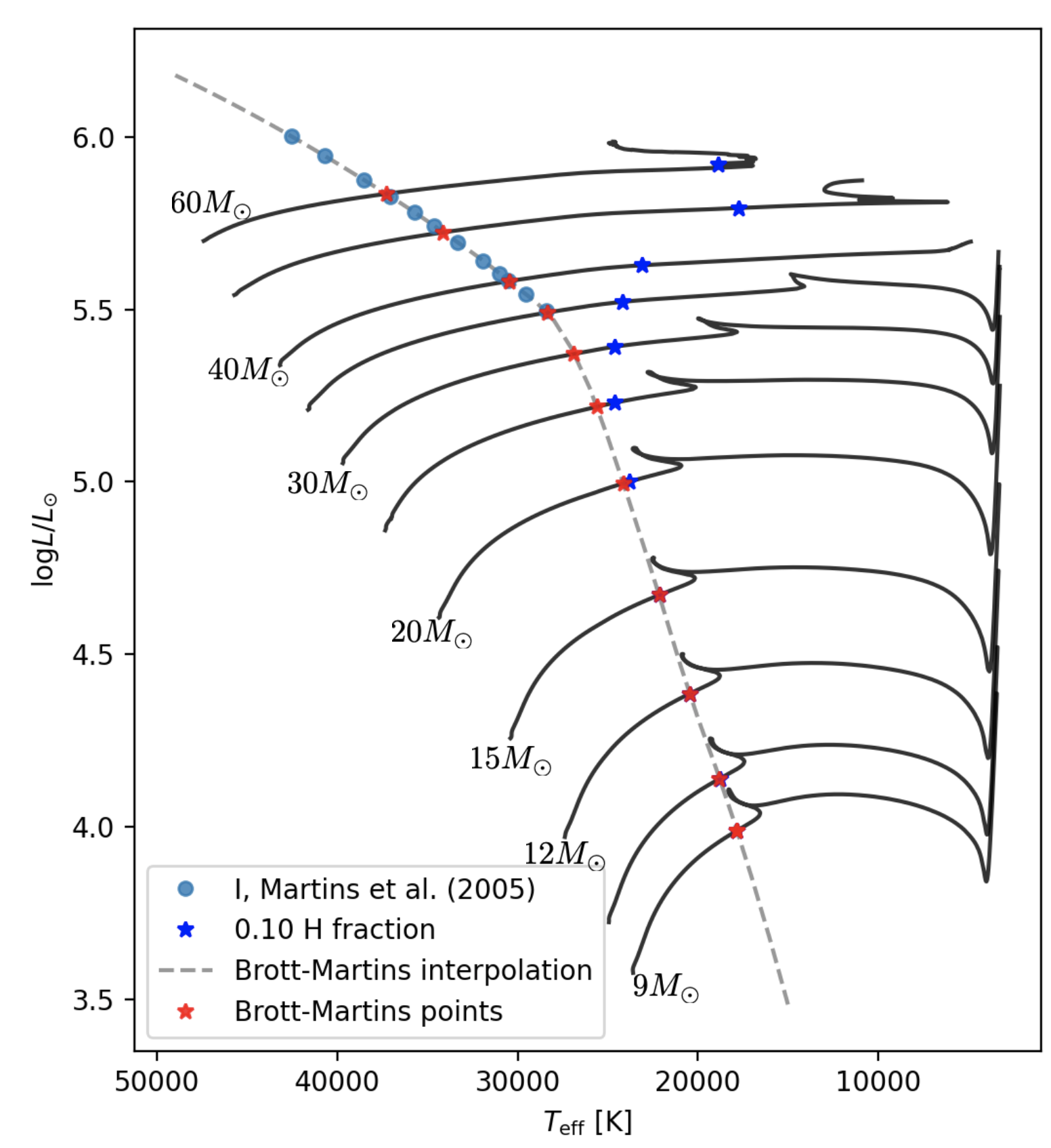}
    \caption{Positions in the HRD for BSGs. Cyan dots are luminosity class I stars from \citet{Martins_2005} and blue stars are the positions on the tracks corresponding to a 10\% core-H fraction. The dashed line represents a second-degree interpolation between the data from \citet{Martins_2005} and the four lowest-mass points with a 10\% core-H fraction (see text for more details).    }
    \label{fig_magnitudes_evolved}
\end{figure}

\subsection{Biases and uncertainties considered in this work}
Stars spend most of their life as main-sequence luminosity-class V (i.e. dwarfs). Main-sequence dwarf stars are hence expected to be the most abundant stars in the Universe. However, when considering a magnitude limited sample, bright stars are typically over-represented (unless the magnitude cut encompasses all stars), especially at larger distances. Non-dwarf stars have a considerable different mass-magnitude relation than dwarf stars and will thus change the AMRF curves. \citetalias{Janssens_2022} only considered the mass-magnitude relation for dwarfs. A consideration of massive binaries with non-dwarf primaries might not only impact the determined AMRF curves, but might also have a large impact on true-negatives and false-positives. In \Secref{section_evolved_primaries}, we investigate the impact of non-dwarf primaries. Here, we also briefly touch upon the impact of systems with a stripped star, which resemble systems such as LB-1 and HR6819.\nextline
In the investigation of false-positives, \citetalias{Janssens_2022} only investigated the contribution of OB+MS binaries. Although triple systems with outer periods shorter than the \Gaias observation time (e.g. 3\,yr for the third data release DR3 or 5\,yr for the fourth data release DR4) are not expected to be abundant \citep[][Tramper et al. in prep.]{Sana_2014}, their possible contribution to the false-positive identification should be investigated, since we are searching for systems that are intrinsically rare. This is done in \Secref{sec_triples_quadruples}, together with a discussion on higher-order hierarchical multiple systems.\nextline
In the ideal scenario, all four parameters in the middle term of \Eqref{eq_AMRF} are known with good accuracy and precision. The astrometric signal $\alpha$, the parallax $\varpi$, and the period $P$ will be published with the astrometric \Gaias orbits. The mass of the primary $M_1$ is for some sources also provided in DR3. However, it is either lacking for most massive stars or inaccurate for most massive stars that have a published $M_1$. Therefore, it must be determined using other data. We discuss the effect of the accuracy of the mass measurement in \Secref{sec_mass_measurements}.\nextline
To determine the fraction of detectable OB+BH systems,  \citetalias{Janssens_2022} required that the astrometric signal produced by a system is larger than three times the \Gaias along-scan variance (see also \Secref{sec_population_of_SGs}). In \Secref{sec_gaia_DR3_results}, we present new predictions on the detectability of OB+BH systems using the basic selection criterion implemented by the \Gaias collaboration in DR3. We also compare these predictions with the ALS\,II sources having an astrometric DR3 orbit.

\begin{figure*}
    \centering
    \includegraphics[width = \linewidth]{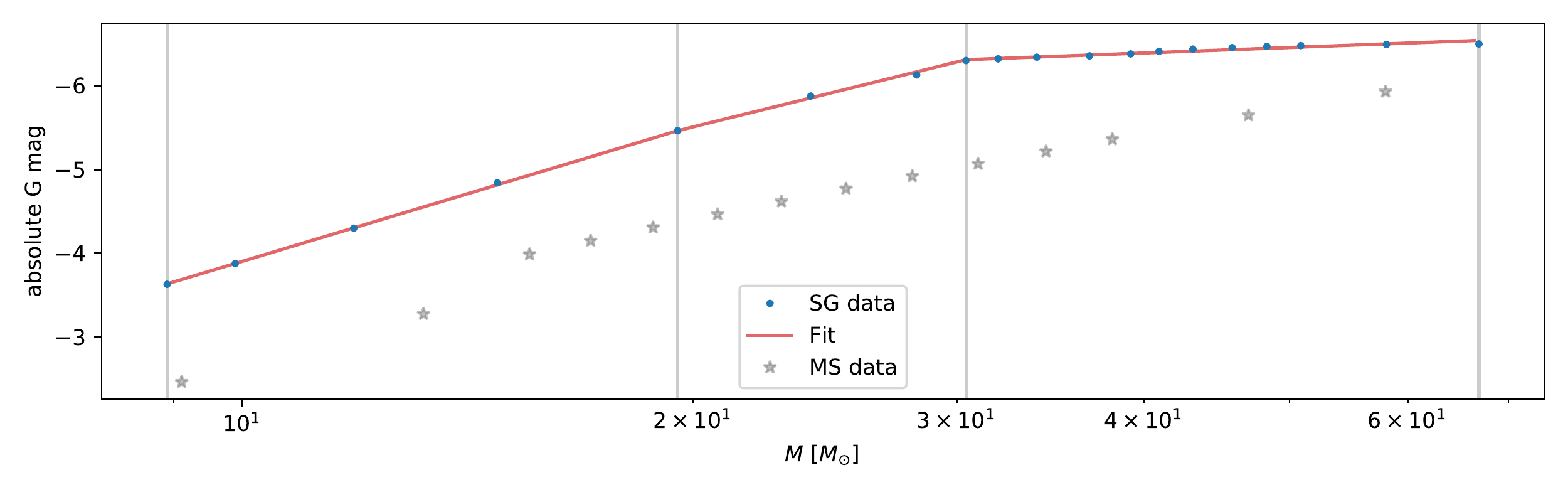}
    \caption{Absolute $G$-band magnitudes for BSGs in the mass range 8.9-66.89\,$\Modot$ obtained from \citet{Martins_2005} and \citet{Brott_2011} (see \Tabref{table_data_mass-mag}). The red solid line shows the fit and the gray vertical lines indicate the different mass regimes. The masses are shown on a logarithmic scale. As a reference, the obtained magnitudes for main-sequence dwarf stars from \citetalias{Janssens_2022} are also shown.}
    \label{fig_mass_magnitude}
\end{figure*}

\section{The impact of an OB supergiant mass-magnitude relation on the AMRF}\label{section_evolved_primaries}
The theoretical AMRF curves are sensitive to the adopted mass-magnitude relation through the band-specific intensity ratio $S$. Evolved non-dwarf primaries or supergiants are substantially brighter than main-sequence dwarf stars. Therefore, non-dwarf primaries can significantly alter the theoretical AMRF curves presented in \Figref{fig_AMRF_curves_SJ22}, potentially altering both the identification fractions as well as the amount of false-positives. \nextline
In order to fully understand which systems might end up as false-positives, systems with evolved supergiant primaries should also be included in the mock population. Since systems with cool red supergiant (RSG) primaries have very long periods ($>5$yrs) due to their large radii, they are not expected to have a binary solution in DR3 or DR4. Potentially, the large convection zones on the surface of RSGs can cause their photocentre to change, creating a chaotic photocentre motion, which is also not in favour of RSGs having a binary solution \citep[][]{Chiavassa_2022}. Therefore, we only considered blue supergiant (BSG) primaries.\nextline 
In \Secref{section_magnitudes_evolved}, we derive a mass-magnitude relation for BSGs appropriate for \Gaia. We obtain the BSG theoretical AMRF curves in \Secref{section_AMRF_curves_SGs}. We simulate a population of binaries with OB and BSG primaries in \Secref{sec_population_of_SGs}, where we also explore the effect of the newly obtained BSG curves. In section \ref{sec_stripped_stars}, we briefly discuss the impact of BH imposters (i.e. stripped He stars) such as the systems LB-1 and HR6819.

\subsection{The G-band magnitudes of BSGs}\label{section_magnitudes_evolved}
In order to create a mock population of BSGs and their theoretical Gaia AMRF curves, the magnitudes of BSGs need to be obtained. As a starting point, we used the parameters derived for stars with a luminosity class I from \citet{Martins_2005}, corresponding to spectral types bewteen O3.5\,I and O9.5\,I and masses $30\Modot \lesssim M \lesssim 66\Modot$. For lower-mass BSGs ($9\Modot \lesssim M \lesssim 30\Modot$), we obtained parameters from the evolutionary tracks published by \citet{Brott_2011} at the time the core-H fraction reaches 10\%. We used the galactic tracks for stars with initial masses in between 9-60$\Modot$ and initial rotational velocities in between 250-300 km s$^{-1}$, corresponding to one track per mass in their model grid. The positions of the BSGs in the Hertzsprung-Russel diagram (HRD) are shown in \Figref{fig_magnitudes_evolved}. \nextline
The data from \citet{Martins_2005} are for O supergiants and does not match the indicated positions for massive stars with a core-H fraction of 10\% on the tracks, where the most massive stars would appear as cool B supergiants. In order to connect both sets of data, we used a second degree spline to interpolate the luminosities between the data from \citet{Martins_2005} and the four lowest-mass points with a core-H fraction of 10\% as a function of effective temperature. This results in the gray dashed line in \Figref{fig_magnitudes_evolved}. The red stars indicate the corresponding position on the evolutionary tracks.\nextline
The same fit was also performed using the first five low-mass points, slightly altering the fit parameters. Hence, the uncertainties in \Tabref{table_G-mag_fit} in the low-mass range are dominated by this effect. Different parameter relations are also plotted in the middle and right panels of \Figref{fig_magnitudes_evolved}. The parameters for the BSGs are given in \Tabref{table_data_mass-mag}.\nextline
The absolute G-band magnitudes of the BSGs were determined in the same way as described in Sec. 3.2 of \citetalias{Janssens_2022}. We derived a log-linear mass-magnitude relation in the same way as described in Sec. 4.1 of \citetalias{Janssens_2022}. The fitted relation is shown in \Figref{fig_mass_magnitude} and the fit parameters are given in \Tabref{table_G-mag_fit}. \nextline
We also explored an additional scenario, using the positions on the HRD tracks where the star has a core-H fraction of 1\%. Equivalents of Figs. \ref{fig_magnitudes_evolved} and \ref{fig_mass_magnitude} are shown in Appendix \ref{appendix_10_vs_1} when using a core-H fraction of 1\% on the tracks (Figs. \ref{fig_magnitudes_evolved_1} and \ref{fig_mass_magnitude_1}). The main difference between the two is for stars with $M<30\Modot$.

\begin{table}
\centering
\caption{Fit parameters for the mass-magnitude relation ($G = a\log (M/\Modot) + b$) of BSGs in different mass regimes.}
\begin{tabu}{ ccc }
     \hline
     \hline
     $M_{\text{low}}-M_{\text{up}}$ [$\Modot$] & $a$ & $b$\\ 
     \hline
     30.41 - 66.89 & $-0.72 \pm 0.01$ & $-5.23 \pm 0.02$\\ 
     8.9 - 30.41 & $-4.89 \pm 0.02$ & $0.96 \pm 0.09$\\
     \hline
     \end{tabu}
\label{table_G-mag_fit}
\end{table}

\begin{figure*}
    \centering
    \begin{subfigure}{0.5\linewidth}
    \includegraphics[width = \textwidth]{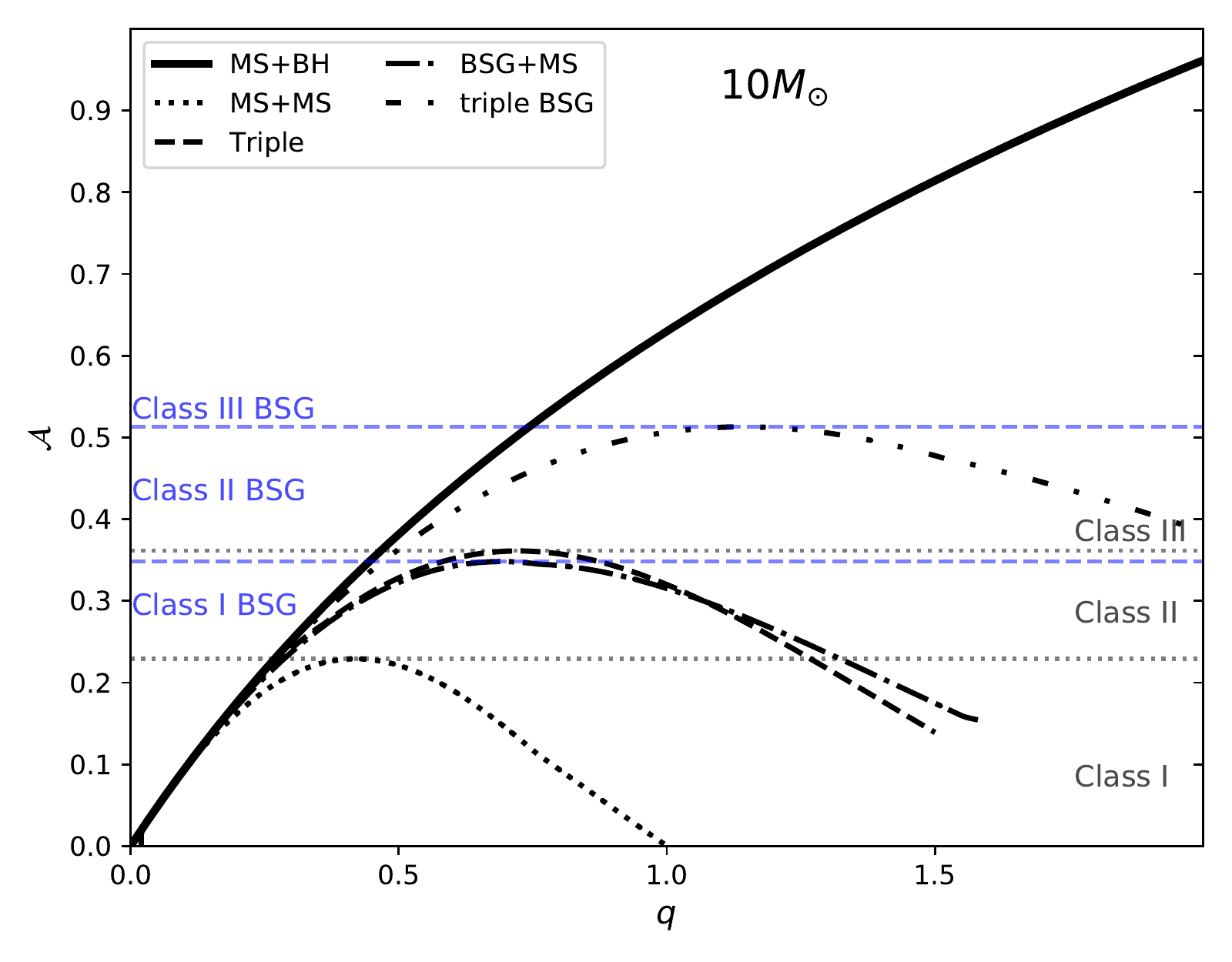}
    \end{subfigure}
    \hspace{-0.1in}
    \begin{subfigure}{0.5\linewidth}
    \includegraphics[width = \textwidth]{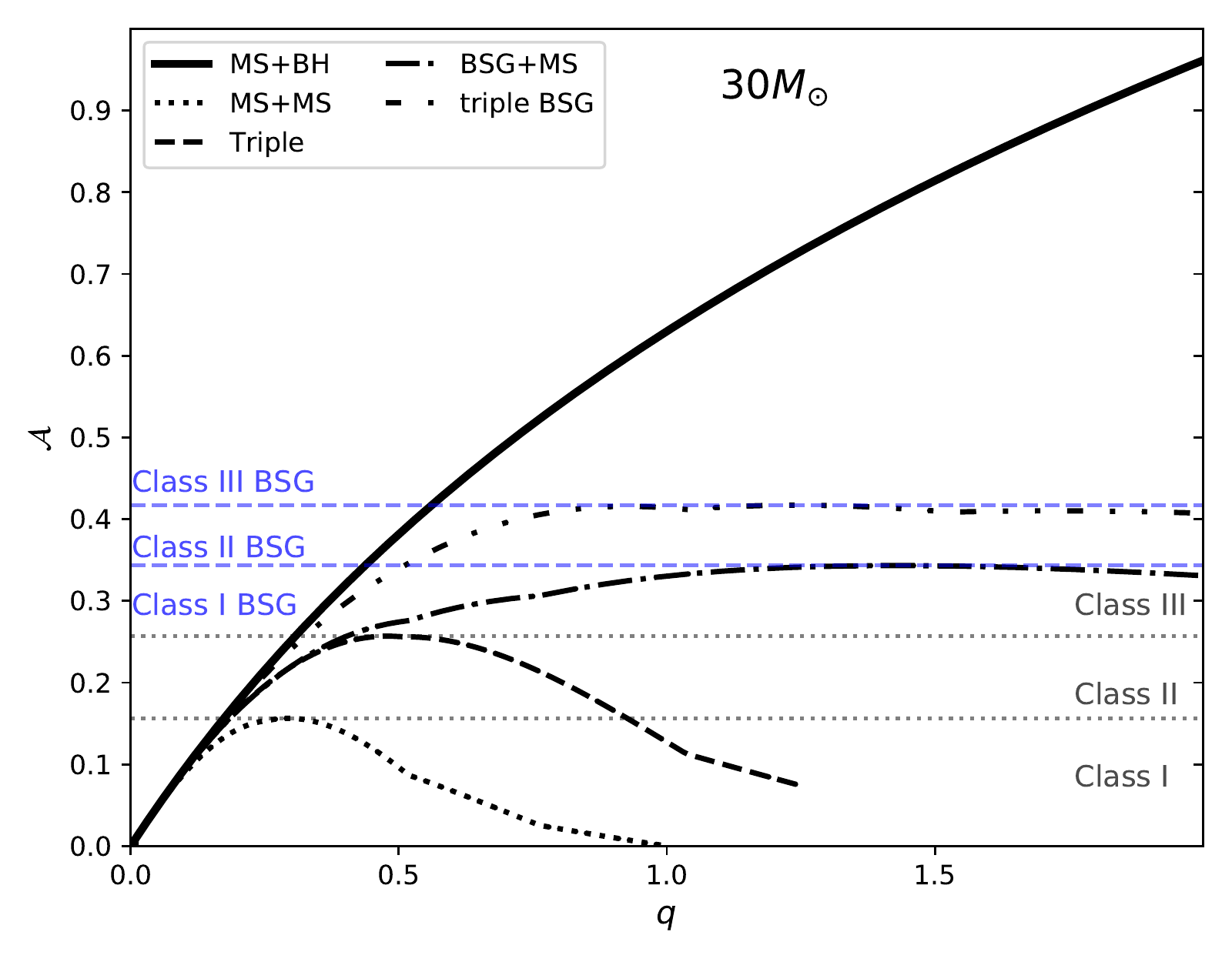}
    \end{subfigure}
    \caption{Different theoretical AMRF curves for two different primary masses of 10\,$\Modot$ (left) and 30\,$\Modot$ (right). In addition to the MS curves, AMRF curves for BSGs are also shown for the same primary masses. The different Classes are now also indicated for both the OBs and the BSGs.}
    \label{fig_amrf_SGs}
\end{figure*}

\subsection{The theoretical AMRF curves for BSGs}\label{section_AMRF_curves_SGs}
Combining the mass-magnitude relation obtained in \Secref{section_magnitudes_evolved} and the right-most term in \Eqref{eq_AMRF}, the theoretical AMRF curves for BSGs can be determined. Unless stated otherwise, the used mass-magnitude relation is that obtained from the interpolation between the data from \citet{Martins_2005} and the four lowest-mass points with a core-H fraction of 10\% on the tracks from \citet{Brott_2011}.\nextline
The BSG curves are shown in \Figref{fig_amrf_SGs} for two different primary masses together with the curves for main-sequence dwarf stars or MS curves. The companion for the BSG-binary curves is assumed to be an unevolved main-sequence dwarf star and for the triple curve a double main-sequence binary companion is assumed. As expected, the BSG curves are reaching higher AMRFs compared to the MS equivalent because BSGs are more luminous and hence yield larger astrometric signals (Eq. \ref{eq_AMRF}). This can also be seen from the horizontal lines, indicating the maximum of the curves and thus the separation between the Classes. For a 10, 20 (not shown), and 30\,$\Modot$ primary, the Class III thresholds increase from 0.36, 0.31, and 0.26 to 0.51, 0.45, and 0.42, respectively. The Class II thresholds change from 0.23, 0.19, and 0.16 to 0.35, 0.30, and 0.34 for a 10, 20 (not shown), and 30\,$\Modot$ primary, respectively.\nextline
To investigate the impact of the adopted core-H fraction, we obtained theoretical AMRF curves using the mass-magnitude relation obtained using a 1\% core-H fraction. A comparison between the 10\% curves and the 1\% curves is shown in \Figref{fig_AMRF_curves_SGs_compare} in Appendix \ref{appendix_10_vs_1}. The 1\% curves can reach values that are even higher than the 10\% curves.

\begin{table*}
    \centering
\caption{Detection, identification, and total identification fractions for different kinds of systems.}
    \begin{tabular}{ c|cccccc }
     \hline
     \hline
      &OB+BH$^{\text{(a)}}$ & BSG+BH$^{\text{(a)}}$ & OB+MS$^{\text{(a)}}$ & BSG+MS$^{\text{(a)}}$ & SMASH+ triples OB$^{\text{(b)}}$ & SMASH+ triples BSG$^{\text{(b)}}$\\ 
     \hline
     $f_{\text{det}}$$^{\text{(c)}}$ [\%]&78$\pm$3&73$\pm$2&22$\pm$1&28$\pm$1&3.4$\pm$0.2&3.7$\pm$0.2\\
     \hline
     MS curves & &  &  &  & & \\
     $f_{\text{id}}$$^{\text{(d)}}$ [\%]&88$\pm$3&87$\pm$3&0.5$\pm$0.2&4.8$\pm$0.7&0.5$^{\text{(f)}}$&0.3$\pm$0.3\\
     $f_{\text{tot}}$$^{\text{(e)}}$ [\%]&68$\pm$3&65$\pm$3&0.12$\pm$0.06&1.4$\pm$0.2&0.01$^{\text{(f)}}$&0.01$\pm$0.01\\
     \hline
      BSG curves & &  &  &  & & \\ 
     
     $f_{\text{id}}$$^{\text{(d)}}$ [\%]&36$\pm$5&40$\pm$4&0.2$\pm$0.2&0.7$\pm$0.3&0.4$^{\text{(f)}}$&0.4$^{\text{(f)}}$\\
     $f_{\text{tot}}$$^{\text{(e)}}$ [\%]&29$\pm$4&29$\pm$3&0.05$\pm$0.04&0.2$\pm$0.1&0.01$^{\text{(f)}}$&0.01$^{\text{(f)}}$\\
     \hline
    \end{tabular}
         \flushleft
    \begin{tablenotes}
      \small
      \item \textbf{Notes.} See Sec. $^{\text{(a)}}$ \ref{sec_population_of_SGs}; $^{\text{(b)}}$ \ref{sec_triple_population} for more details on the simulations; $^{\text{(c)}}$ the fraction of systems detectable by \Gaia; $^{\text{(d)}}$ the fraction of detectable systems that are identifiable with the AMRF; $^{\text{(e)}}$ the fraction of identifiable systems in the full simulated population ($f_{\text{det}} \times f_{\text{id}}$); $^{\text{(f)}}$ the values represent the maximum and not the average
    \end{tablenotes}
\label{table_fractions}
\end{table*}

\begin{table*}
    \centering
\caption{Estimated number of sources detected and identified.}
    \begin{tabular}{ c|cccccc }
     \hline
     \hline
      &OB+BH$^{\text{(a)}}$ & BSG+BH$^{\text{(a)}}$ & OB+MS$^{\text{(a)}}$ & BSG+MS$^{\text{(a)}}$ & SMASH+ triples OB$^{\text{(b)}}$ & SMASH+ triples BSG$^{\text{(b)}}$\\ 
     \hline
     detected&117&110&770&980&221&259\\
     \hline
     MS curves & &  &  &  & & \\
     identified&102&95&4&48&1$^{\text{(c)}}$&1\\\hline
     BSG curves & &  &  &  & & \\
     identified&42&43&1&7&1$^{\text{(c)}}$&1\\
     \hline
    \end{tabular}
         \flushleft
    \begin{tablenotes}
      \small
      \item \textbf{Notes.} See Sec. $^{\text{(a)}}$ \ref{sec_population_of_SGs}; $^{\text{(b)}}$ \ref{sec_triple_population} for more details on the simulations; Numbers are estimated using the central or maximum ($^{\text{(c)}}$) fractions in \Tabref{table_fractions}.
    \end{tablenotes}
\label{table_numbers}
\end{table*}

\subsection{A mock population including BSG primaries}\label{sec_population_of_SGs}
We have performed a cross-match between the ALS\,II and the Astronomical database SIMBAD\footnote{\url{http://simbad.cds.unistra.fr/simbad/}} to obtain spectral types for the sources in the ALS\,II. Of those that have a spectral type, about 50\% had at least one companion which was not a dwarf (luminosity class V) and hence more evolved. With this information, we simulated a population of binaries where 50\% of the systems have a BSG primary.\nextline
We performed a Monte Carlo (MC) simulation to immediately include the effect of measurement uncertainties. Following \citetalias{Janssens_2022}, we take the uncertainties on $\alpha$ and $\varpi$ equal to the derived along-scan variance data provided to us by A. Everall \citep[priv. comm., based on a similar method described in][]{Everall_2021}. We use a 10\% uncertainty on $P$ and a 50\% uncertainty on $M_1$. The published DR3 periods have on average a 3\% precision, smaller than what is used here. However, in order to directly compare to the predictions of \citetalias{Janssens_2022}, a 10\% uncertainty on the period is used. If the precision on $P$ would be lowered, this could result in a slightly lower fraction of false-positives, as it decreases the mobility of the data points in the AMRF diagram. For the systems, $\Omega$ (longitude of the ascending node) and $\omega$ (argument of periastron) were uniformly randomly distributed between $0$ and $2\pi$, and $\cos i$ (where $i$ is the inclination) between $-1$ and $1$. Distances were drawn from the distribution of distances in the ALS\,II.\nextline 
In total we simulated 1000 samples of 150 OB+BH, 150 BSG+BH, 3500 OB+MS, and 3500 BSG+MS binaries. The numbers originate from the estimations done in Sec. 6.3 of \citetalias{Janssens_2022}. From the $>13\,000$ sources in the ALS\,II, 50\% would be post-interaction binaries \citep[][]{De_Mink_2014}, amounting to $\sim$7000 sources. Furthermore, 3\% of OB stars in binaries are expected to have a BH companion \citep[][]{Langer_2020}, resulting in almost 300 OB+BH systems. Using now that 50\% of sources in the ALS\,II are non-dwarf stars, we obtain the above-stated population.\nextline 
To determine which systems are detectable by \Gaia, we adopted the same assumptions as \citetalias{Janssens_2022} used for DR3: $P<3$\,yr and $\alpha>3\sigma_{G,\text{DR3}}$. We note that, due to $\alpha$ always being positive, the criterion of $\alpha>3\sigma_{G,\text{DR3}}$ may result in a larger fraction of single stars falsely identified as binaries in comparison to that expected  in the case of a normal distribution (0.3\%). This criterion is hence an optimistic one and, as discussed in \Secref{sec_gaia_DR3_results}, it is currently not used for DR3. However, we do adopt it here to facilitate a comparative study to \citetalias{Janssens_2022}.\nextline
Also similar to \citetalias{Janssens_2022}, identifiable systems need to be detectable and have $\mathcal{A}-\sigma_{\mathcal{A}} > \mathcal{A}_{\text{triple}}$. Here, $\mathcal{A}_{\text{triple}}$ can be either from the MS or the BSG curves. Unless stated otherwise, these assumptions will hold for the remainder of the manuscript. \nextline 
The average detection and identification fractions are given in the first four columns of \Tabref{table_fractions} and the estimated numbers of sources in the ALS\,II are given in \Tabref{table_numbers}. In \Tabref{table_fractions}, $f_{\text{det}}$ is the fraction of systems detectable by \Gaias, $f_{\text{id}}$ is the fraction of the systems detected by \Gaias that would be identified with the AMRF method, and $f_{\text{tot}}$ is the fraction of systems identified in the full simulated population, defined as
\begin{equation}\label{eq_total_fraction}
    f_{\text{tot}} = f_{\text{det}} \times f_{\text{id}}.
\end{equation} 
\nextline
First, we can see that the usage of the BSG curves greatly reduces the fraction of identifiable OB/BSG+BH systems -- true-positives -- by a little more than half. For the OB+BH and BSG+BH systems, the identification fractions now decrease to 29\% compared to 68\% and 65\%, respectively, when using the MS curves. Assuming around 300 OB/BSG+BH systems, about 85 BHs can be detected compared to almost 200 when using the MS curves. Hence, a significant fraction of the population is lost when using the BSG curves.\nextline
Second, using the BSG curves also reduces the amount of false-positives. When using the MS curves, the fraction of false-positive OB+MS binaries is around 0.1\% and it is well below 0.1\% when using the BSG curves. With an estimated 3500 of OB+MS systems in the ALS\,II, the number of false-positive systems would be between $\sim$1-6 depending on the used curves. This is indeed a negligible amount just as shown in \citetalias{Janssens_2022}.\nextline
The false-positive fraction of BSG+MS systems is on average about 1.3\% when using the MS curves. Given 3500 systems, this would translate to $\sim$50 systems in the ALS\,II. With an estimated 200 OB/BSG+BH systems, this would mean 20-25\% of the identified sources are false-positive. When using the BSG curves, this fraction decreases to 0.2\%, which would result in $\sim$7 systems compared to $\sim$85 OB/BSG+BH binaries.\nextline
We thus see that the inclusion of BSGs changes the identification fractions. Ideally, the evolutionary stage of the primary is known and each system can then be compared to the curves corresponding to the evolutionary stage of the primary to benefit both from the high identification fraction of OB+BHs and the low false-positive fraction of BSG+MS systems. However, if the evolutionary stage of the primary is not known one might consider using the BSG curves to obtain the least amount of false-positives despite the significant decrease in true-positives.

\subsection{BH imposters like LB-1 and HR~6819}\label{sec_stripped_stars}
The systems LB-1 \citep[][]{Liu_2019} and HR~6819 \citep[][]{Rivinius_2020} are two of the most recent and well known BH imposters. Both these systems host a Be star bound to a bloated stripped star that is 5-10 times less massive than its Be companion, resulting in an extreme mass ratio. Such binaries are in an extremely rare evolutionary phase, caught as the stripped star contracts towards the helium main-sequence, while it is still extended and visually bright. The fact that the stripped star is overluminous compared to its mass can make these systems appear as OB+BHs in the AMRF diagram. However, according to \citet{Bodensteiner_2020}, \citet{Shenar_2020}, and \citet{El_badry_2020}, LB-1 and HR~6819  consist of two components with very similar visual brightness. This hence means that the visual photocentre of these systems is located very close to the centre of mass and thus the photocentre motion is very small. Therefore, we would not expect these systems to be astrometrically detected as binaries in DR3. 

Nevertheless, it is possible that other "LB-1"-like systems exist in which the stripped star is substantially brighter than its more massive companion. In such cases, the system could be erroneously identified as an OB+BH binary using our method. Given the overall rarity of such binaries and the associated uncertainties of binary evolution, it is difficult to assess the contamination of such systems among the OB+BH candidates. However, a rough estimate can be made using the prototypical example of HR~6819. This is the only binary within a $\approx 350\,$pc volume that is known to host a bloated stripped star. There are roughly 300 OB stars in the ALS~II catalogue within this volume, and, according to \citet{Langer_2020}, $\approx 2$\% should host BHs, amounting to roughly six targets. Hence, the population of OB+BH might be expected to be larger by a factor of a few. This factor increases if one considers that the bloated stripped stars need to be significantly more luminous than their companions to be confused as OB+BH binaries astrometrically. An important caveat here, however, is that the estimated number of OB+BH binaries is a prediction, not an observation, while the estimated number of binaries hosting bloated stripped stars hinges on the assumption that HR~6819 is unique within this volume. 

\citet{El-Badry_2022_SB1s} analysed a sub-sample of single-lined spectroscopic binaries (SB1) detected by \Gaias in DR3 and reported they all contain a bloated stripped star. While no \Gaias SB1 compact-object systems have been reported so far, one astrometric BH has been reported \citep[][]{El-Badry_2022_SunBH}, orbiting a low-mass star. This could indicate that bloated stripped stars are more common than BHs. However, the stripped stars around massive stars most likely originate from another massive star, whereas for the low-mass binaries detected in DR3, the stripped stars most likely originate from other low-mass stars due to the selection biases of DR3. Thus, the stripped-star and BH populations in these DR3 low-mass binaries originate from different progenitors. In contrast, for massive stars, both stripped stars and BHs have the same progenitors. Hence, the apparent overabundance of stripped stars in low-mass binaries in DR3 compared to BHs cannot be directly extended to the massive star domain.

To conclude, the numbers currently indicate that, among a population of OB+BH candidates, not more than 10-20\% should be "LB-1"-like systems erroneously classified as OB+BH. While relatively small, this number is not negligible. Moreover, it is possible that the true proportion of such imposters is larger (e.g. if these systems are more common, or if OB+BH binaries are rarer than predicted). Dedicated spectroscopic and interferometric surveys of the candidates would be needed to fully remove this uncertainty \citep[e.g.][]{El-Badry_2022_SB1s,Frost_2022, Shenar_2020}.\nextline

\section{Higher-order multiple systems}\label{sec_triples_quadruples}
The theoretical AMRF curves for OB+MS binaries cover much lower AMRF values than those for triples. Indeed, triple systems can produce larger astrometric signals, as a binary companion is less bright than a main-sequence star of equal mass. Observational errors thus can more easily cause a triple system to become a false-positive than for an OB+MS binary. Here, we investigate the impact of triple systems and higher-order multiple systems on the amount of false-positives.\nextline
The number of hierarchical triples or higher-order multiple systems that have a period shorter than $\sim$3-5 years are limited \citep[$\sim$10\%;][]{Sana_2014}. On top of that, the binary orbits published in DR3 satisfy hard quality cuts (see \Secref{sec_parallax_precision} for some details). Chaotic photocentre motions caused by higher-order multiple systems where all three components significantly contribute to the light will thus most likely not have a clear astrometric solution even though for DR4 the quality cuts will probably be slightly less conservative than those for DR3. Therefore, based on the \Gaias detection alone, we do not expect higher-order multiple systems to greatly impact the amount of false-positives. However, there might be systems with outer periods $P\lesssim3$\,yr where the photocentre motion is not extremely chaotic that will have an astrometric binary solution and the impact of these needs to be investigated.\nextline
First, we explored the effect of different triple configurations on the theoretical AMRF curves (\Secref{sec_kind_of_triples}). Then, in \Secref{sec_triple_grid}, we simulated a grid of triples to investigate which triple configuration has the largest impact on the amount of false-positives. Next, in \Secref{sec_triple_population}, we used the determined distributions for the triples in the Southern MAssive Stars at High angular resolution survey \citep[SMASH+][]{Sana_2014} to simulate a population of triples. Last, we discuss the impact of quadruples and higher-order multiple systems in \Secref{sec_higher_order}.

\subsection{Investigation of distinct triple configurations}\label{sec_kind_of_triples}
\begin{figure}
    \centering
    \includegraphics[width = \linewidth]{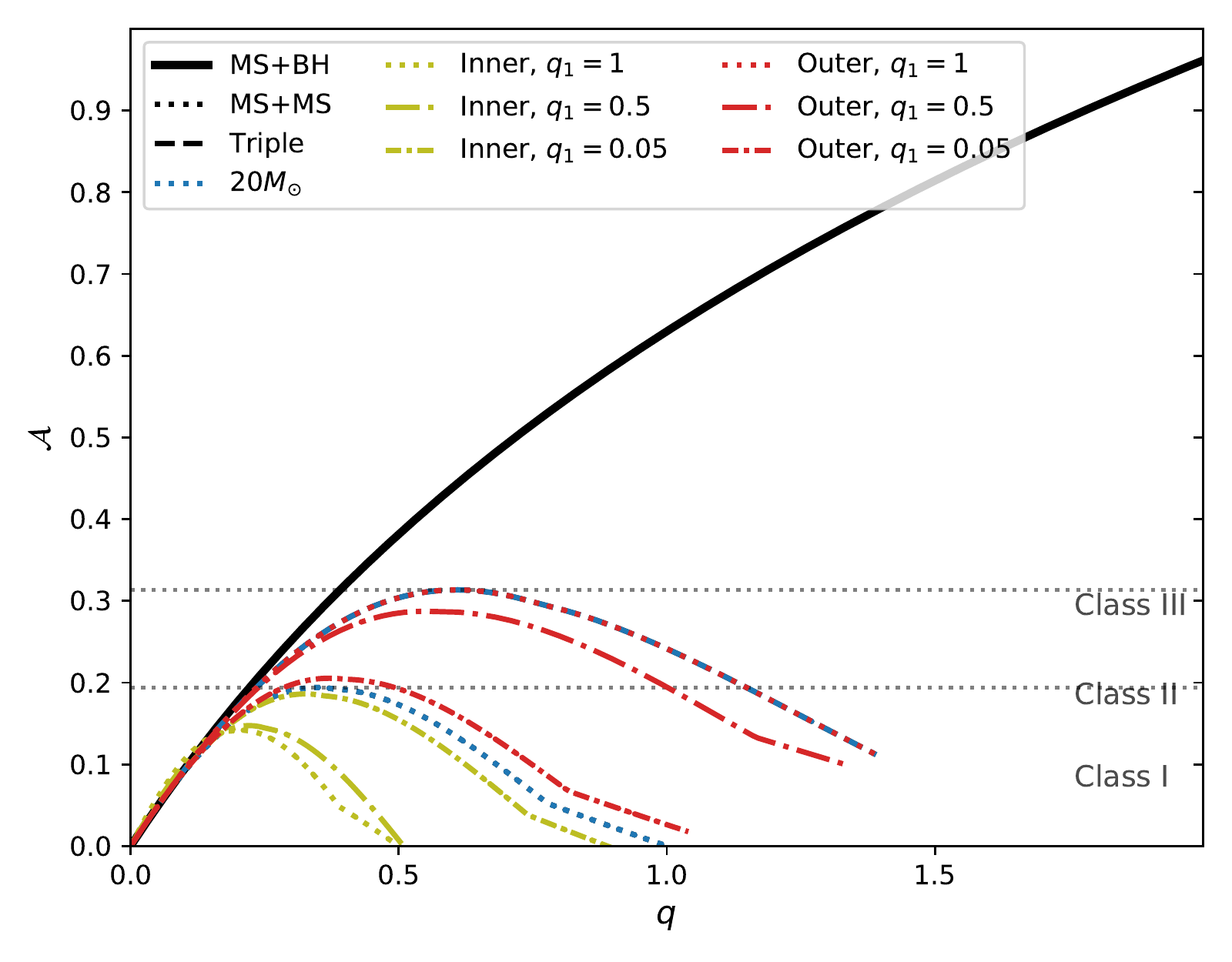}
    \caption{Different theoretical AMRF curves for a 20\,$\Modot$ primary. Triple curves for different mass ratios of the inner binary $q_1$ are shown for the case where the inner binary is the most luminous (`Inner') or the outer binary is more luminous (`Outer').}
    \label{fig_amrf_triples_empty}
\end{figure}

The theoretical AMRF curves for triples in \Figref{fig_AMRF_curves_SJ22} represent the case where the outer star is more luminous than the inner binary. On top of that, the curves also assume that the inner binary has a mass ratio of unity ($q_1 = 1$, not to be confused with the $q$ on the x-axis, which is in this case the mass of the inner binary divided by that of the outer star). Only the curves for $q_1 = 1$ are shown because they are the `worst case' scenario (i.e. they give the highest AMRF curves). Starting at $q_1=1$, which coincides with the previously presented curves (\Figref{fig_AMRF_curves_SJ22}), the AMRF curves keep on moving downward for decreasing $q_1$, with the limit being a binary system at $q_1 = 0$ (\Figref{fig_amrf_triples_empty}). The inner binary appears underluminous compared to a single star with a mass equal to the total mass of the inner binary. Hence, the photocentre will be located closer to the outer star and produce a larger astrometric signal for any $0<q_1 \leq 1$ than compared to a binary system.\nextline
What is thought to be more common are triples where the most luminous component is located in the inner binary. For these systems, $q$ is the mass of the outer star divided by that of the inner binary and $M_1$ in \Eqref{eq_AMRF} is the mass of the most luminous star in the binary. We can distinguish the two extremes: $q_1 = 1$ and $q_1 = 0$. The latter case again reduces the system to a binary system. In contrast to the other kind of triples, where $q_1 = 1$ is the `worst case' scenario, here $q_1 = 1$ is the `best case' scenario, giving the lowest AMRF curves (\Figref{fig_amrf_triples_empty}). Following the same reasoning as before, the inner binary looks underluminous compared to a single star with a mass equal to the total mass of the inner binary. Hence, the photocentre will be located farther away from the inner binary and, in this case, produce a smaller astrometric signal for any $0<q_1 \leq 1$ than compared to a binary system.\nextline
From the theoretical AMRF curves presented in \Figref{fig_amrf_triples_empty}, we can argue that triples where the most luminous component is the inner binary will not contribute significantly to the amount of false-positives. The reason for this is that these kind of triples are all falling below the OB+MS curves and that OB+MS binaries are already giving very few false-positives (\citetalias{Janssens_2022} and \Secref{sec_population_of_SGs}). Thus, triple systems where the outer star is the most luminous are the only triples with the potential to cause many false-positives, as these have AMRF curves above those of OB+MS binaries.

\begin{figure*}
    \centering
    \begin{subfigure}{\linewidth}
    \includegraphics[width = \textwidth]{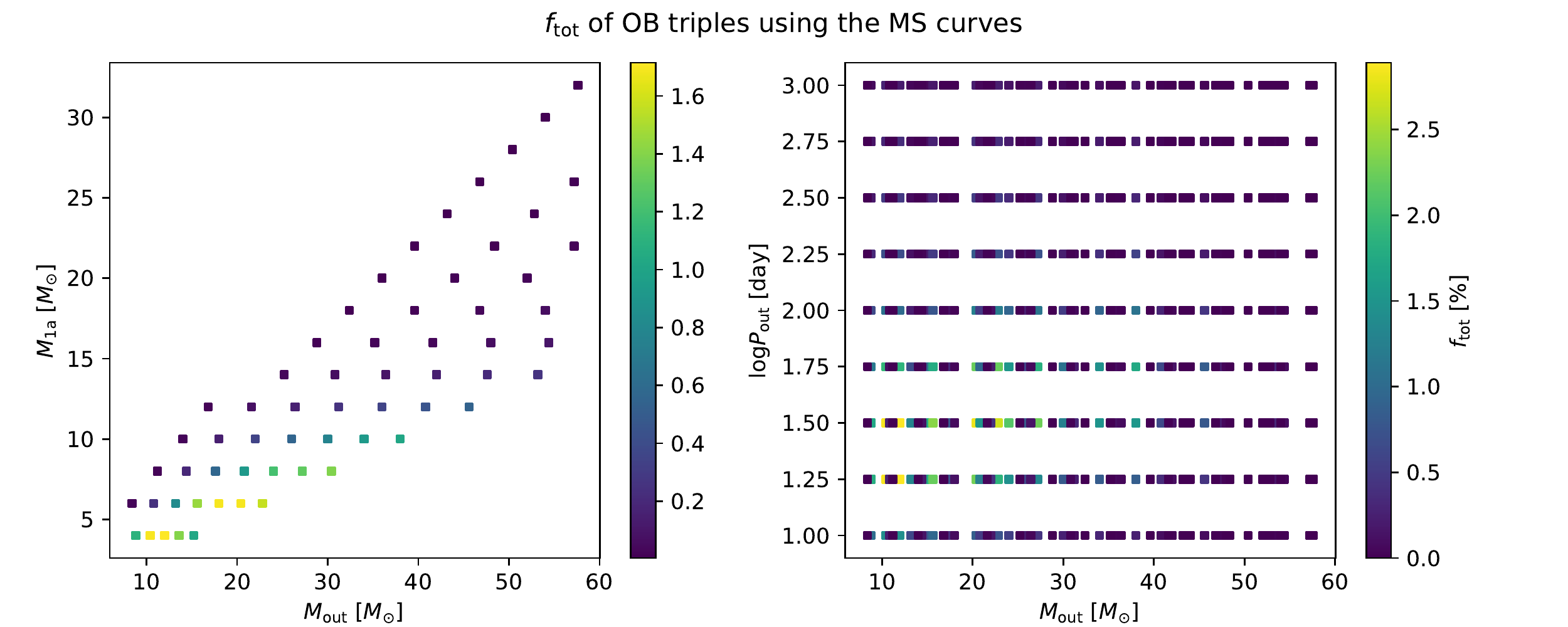}
    \end{subfigure}
    \begin{subfigure}{\linewidth}
    \includegraphics[width = \textwidth]{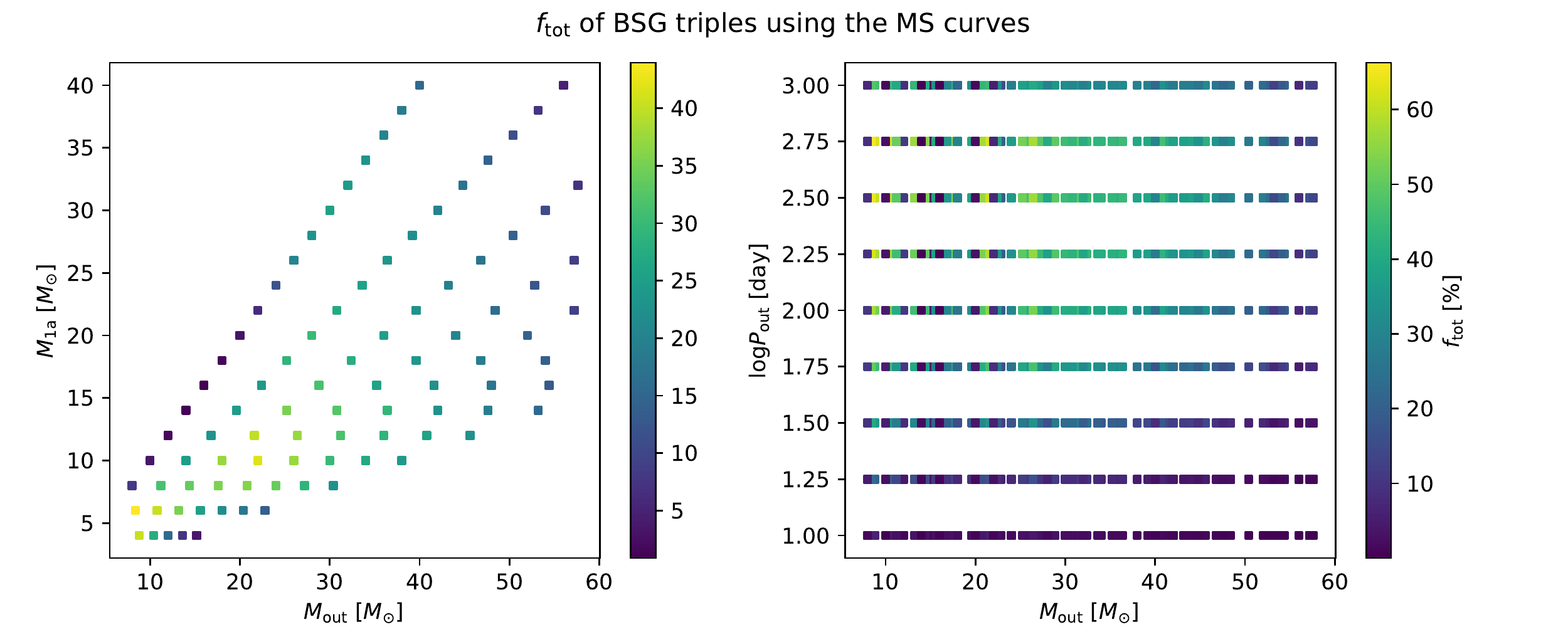}
    \end{subfigure}
    \caption{Total identification fraction for triple systems where the outer star is the most luminous component and the inner binary has $q_1 = 1$. In the left panels, $M_{\text{1a}}$ is the mass of one of the components in the inner binary. The top panels are for OB systems, the bottom panels are for systems with a BSG outer star.}
    \label{fig_grid_identified_part_OB}
\end{figure*}

\subsection{A grid approach}\label{sec_triple_grid}
\begin{figure*}
    \centering
    \begin{subfigure}{\linewidth}
    \includegraphics[width = \textwidth]{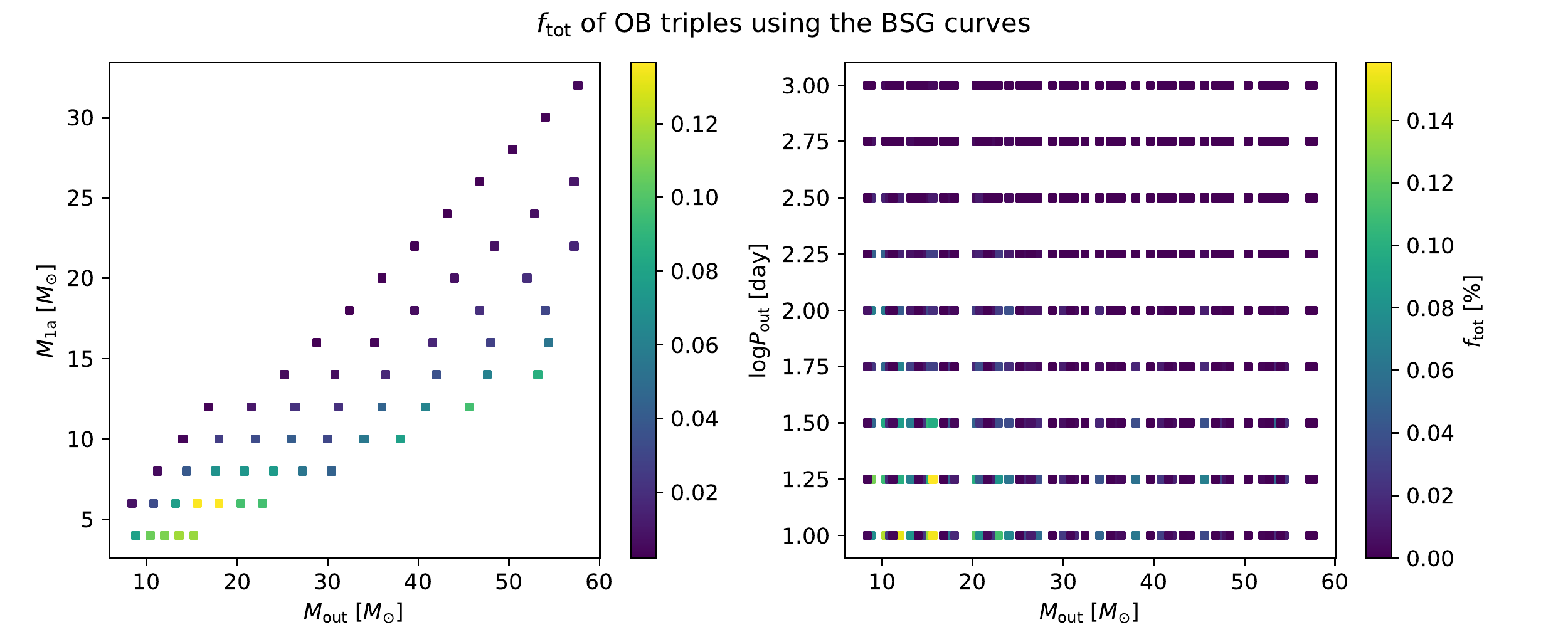}
    \end{subfigure}
    \begin{subfigure}{\linewidth}
    \includegraphics[width = \textwidth]{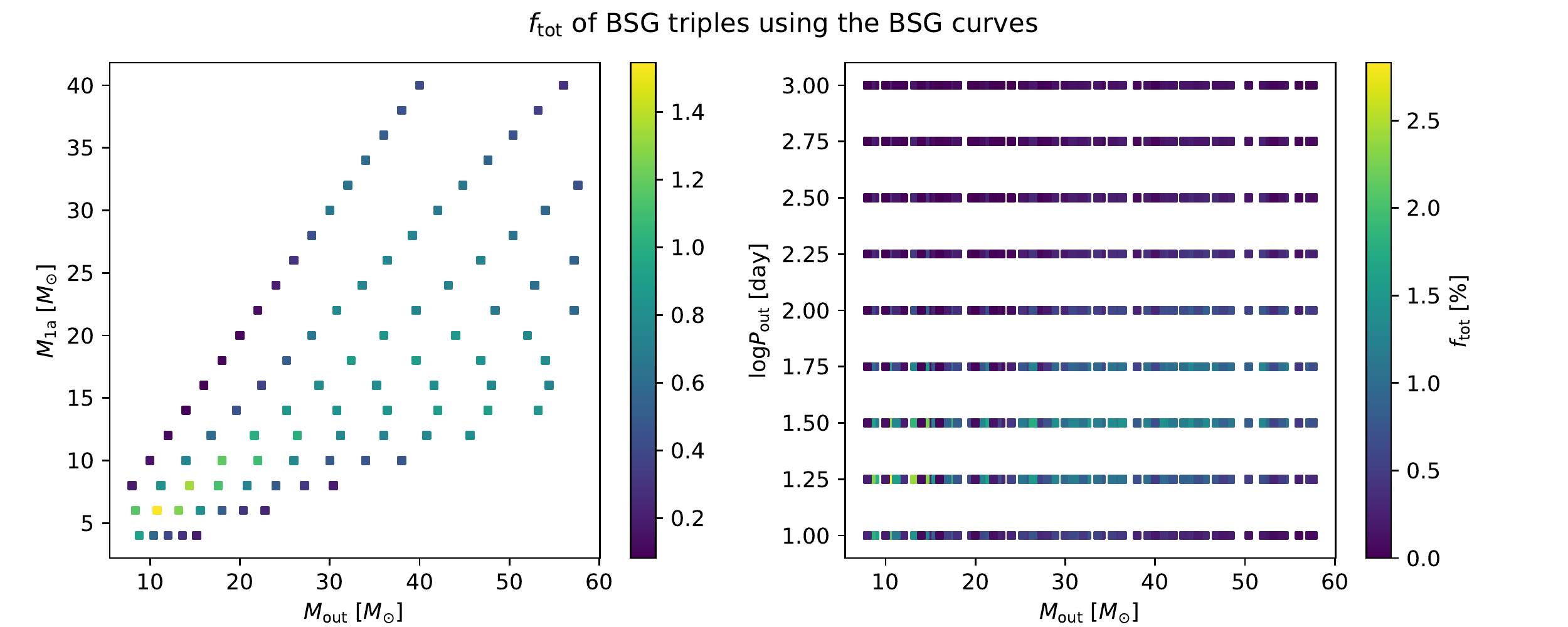}
    \end{subfigure}
    \caption{Same as \Figref{fig_grid_identified_part_OB}, but when using the BSG curves.}
    \label{fig_grid_identified_part_SG}
\end{figure*}

As previously discussed, the worst-case scenario is that in which the binary companion has a mass ratio equal to unity. Therefore, we only simulated triples with $q_1 = 1$. We varied the outer eccentricity between 0 and 0.75 with a step size of 0.25. The outer period $P_{\text{out}}$ ranged from 10 to 1000 days with a logarithmic step size of 0.25. For the masses, we varied the mass of each component in the inner binary between 4 to 40 $\Modot$. We varied the mass ratio $q$ of the triple system from 0.1 to 2.1 with a step size of 0.2, such that the mass of the outer star is equal to $M_{\text{out}} = 2M_1q$. We put a lower and upper limit on $M_{\text{out}}$, which are 8\,$\Modot$ and 60\,$\Modot$, respectively.\nextline
Again, we performed a MC simulation with the same uncertainties and randomly drawn orientations and distances as described in \Secref{sec_population_of_SGs}. For each combination of the above-stated parameters, 6000 systems were simulated with random orientations and distances. \nextline
The total identification fractions (i.e. the fraction of systems detected by \Gaias and identified with the AMRF) are shown in \Figref{fig_grid_identified_part_OB} when using the MS curves, for the mass of one of the components of the inner binary $M_{\text{1a}}$ ($M_{\text{1a}} = M_{\text{1b}}$) or the outer period $P_{\text{out}}$ as a function of the mass of the outer star $M_{\text{out}}$. More parameters are shown in \Figref{fig_grid_OBs_full}. The fraction of false-positive triples for OB stars is fairly low ($<3$\%), even at the grid points giving the highest false-positive fraction. However, for triple systems where the outer star is a BSG, the false-positive identification fraction can reach values of 50\%, which is extremely high. This however does not result from the fact that the system is triple but because the AMRF curve for dwarfs are inapropriate for BSGs (\Secref{section_evolved_primaries}).\nextline
In \Figref{fig_grid_identified_part_SG}, the same false-positive fractions are shown but now when using the BSG curves (and more parameters in \Figref{fig_grid_SGs_full}). This decreases the false-positive fraction among systems with BSG primaries to $<3$\%. Hence, the usage of the BSG curves can significantly reduce the amount of false-positives in case the luminosity class of the primary is not known. However, again, the usage of the BSG curves also significantly reduces the amount of true-positives. Therefore, it is highly recommended and beneficial to establish the nature of the most luminous component to still obtain the high amount of true-positives among the OB+BH systems.

\subsection{A mock population of triples}\label{sec_triple_population}
The SMASH+ has observed a significant amount of triple systems with massive star components. The SMASH+ combines high angular resolution data of massive stars in the southern hemisphere with (most often spectroscopic) information found in the literature to investigate the multiplicity of massive stars over the full range of separation up to $>10\,000$\,au. Tramper et al. (in prep.) have derived masses and physical separations for the companions detected with interferometry. Using this information, we can determine distributions for the periods and mass ratios of the triple systems.\nextline
Practically all of the triples in the SMASH+ database have an inner binary which is more luminous compared to an outer third component. As explained in \Secref{sec_kind_of_triples}, triples where the inner binary is more luminous theoretically have a smaller $\mathcal{A}$ than binaries. Hence, we do not expect those to significantly contribute to the amount of false-positives. To verify this, we have simulated a population of triples using the distributions derived from SMASH+ (Tramper et al. in prep.). While doing this, we have assumed that the photocentre motion of the inner binary combined with the photocentre motion of the triple is not chaotic, but resembles a binary (i.e. that the point-spread function of the triple would not be impacted by the inner-binary orbit). \nextline
For triples where the outer star is the most luminous, it is effectively more difficult to identify those as triples instead of binaries. Therefore, the SMASH+ database might be biased against such systems though a few of these are observed. Nonetheless, such triples are not expected to be very abundant.\nextline 
We simulated both triples with all main-sequence dwarf components and triples where the most luminous star is a BSG. A similar approach was used as described in \Secref{sec_population_of_SGs}, where we used a MC simulation with similar uncertainties and random orientations and distances. In total, 1000 samples of 7000 systems were simulated both for OB and BSG primaries.\nextline
The detection and identification fractions can be found in the two last columns of \Tabref{table_fractions} and the number of estimated sources in \Tabref{table_numbers}. The fraction of triples estimated to be detected by \Gaias DR3 is only $\lesssim$4\%. This is mostly because either the outer period of the system is much larger than the DR3 baseline or because the mass ratio of the outer star to the inner binary system is very low, resulting in a very small photocentre motion.\nextline
As expected, the false-positive identification fraction of the SMASH+ triples detected in DR3 is extremely low (around 0.01\%), resulting in potentially one false detection. Since most triples have fairly large mass ratios for the inner binary, these systems will produce much smaller photocentre motions as binaries and end up at lower values for $\mathcal{A}$, making it more difficult for these systems to be falsely-identified. If instead we would use the BSG curves, the identification fractions become practically zero. Unlike with binaries, for the triple population as observed by the SMASH+, there is no direct need to use BSG curves instead of MS curves.

\subsection{Quadruples and higher-order multiple systems}\label{sec_higher_order}
Besides triples, higher-order multiple systems exist and contribute to $\sim$5-10\% of the sample studied by SMASH+. There are two kinds of hierarchical quadruple systems: (i) two binaries orbiting around each other \citep[e.g. QZ Car][]{Rainot_2020} and (ii) hierarchical systems formed by a binary orbited by two other stars at increasing separation \citep[e.g. Plaskett's star][]{Linder_2008,Sana_2014}. In both cases, the binaries themselves most likely will be unresolved. In the hierarchical case, the photocentre motion will be very chaotic, unless one or both of the outer stars are very low mass and thus have a relatively low luminousity compared to the inner binary. If only one star is very low mass, the photocentre motion approximately reduces to that of a triple system, a scenario which is already discussed. In case both outer stars are low-mass stars, the photocentre motion will be very small, because the luminous binary in the quadruple will not show large motion. Hence, it is likely that these sources will not have an astrometric binary solution in DR3 or DR4. Alternatively, the published astrometric solution is that of the inner binary, in which case the quadruples should not greatly impact the amount of false-positives.\nextline
For the quadruples in case (i), there are three (extreme) configurations where the photocentre motion will resemble closely that of a binary system : (a) two binaries each with a mass ratio equal to unity (e.g. (15+15)$\Modot$ + (10+10)$\Modot$), (b) one binary has an extreme mass ratio (e.g. (29+1)$\Modot$ + (10+10)$\Modot$), and (c) both binaries have extreme mass ratios (e.g. (29+1)$\Modot$ + (19+1)$\Modot$).\nextline
An (a)-type quadruple gives the same AMRF as a binary and is most easily explained by an example. In the example of a (15+15)$\Modot$ + (10+10)$\Modot$, this quadruple has the same mass ratio and intensity ratio as a (15+10)$\Modot$ binary. Hence, the theoretical AMRF reduces to that of a binary. As shown in \citetalias{Janssens_2022}, binaries are not expected to contribute to the false-positives and thus neither will these kind of quadruples.\nextline
The case (b) quadruple resembles very closely a triple system. As shown in Secs. \ref{sec_triple_population} and \ref{sec_triple_grid}, the number of false-positives expected from triples is very low and will even be lower for these kind of quadruples as they are even less abundant.\nextline
The last case, case (c), resembles closely the case of a binary system. Due to the extreme mass ratios of both binary components, the AMRF of these kind of systems will approach that of binaries. Hence, we also do not expect case (c) quadruples to affect the false-positives.\nextline
For even higher-order multiple systems, like quintuples, the reasoning is very similar to that of quadruples. Either the photocentre motion is very chaotic, and no solution for this system will be given in DR3, or the systems photocentre motion can be approximated by any of the previous discussed cases (binaries, triples, and quadruples). Hence, higher-order multiple systems should also not impact the false-positive OB+BH identifications.

\begin{figure*}
    \centering
    \begin{subfigure}{\linewidth}
    \includegraphics[width = \textwidth,trim={0 0.95cm 0 0},clip]{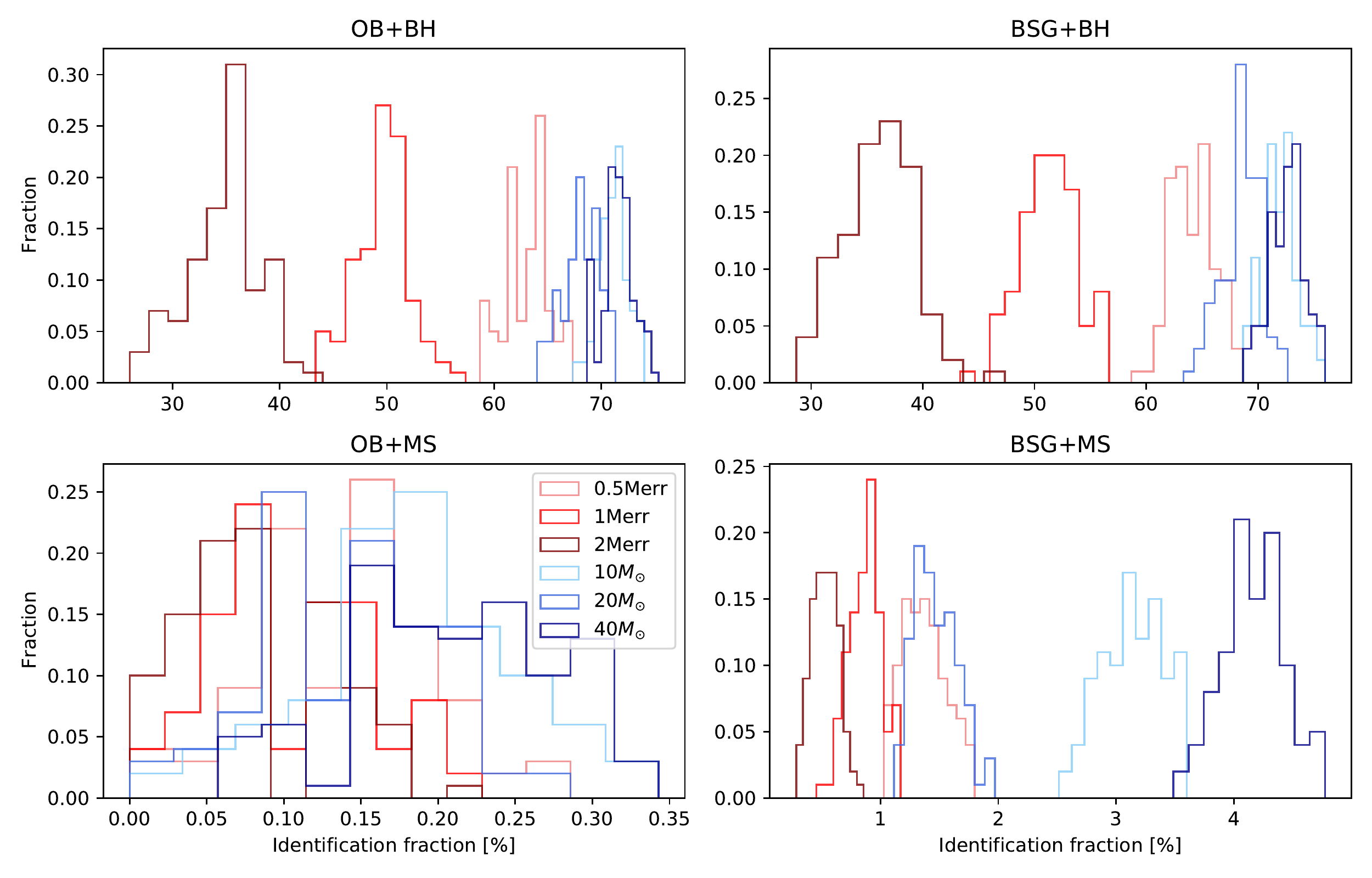}
    \end{subfigure}
    \begin{subfigure}{\linewidth}
    \includegraphics[width = \textwidth]{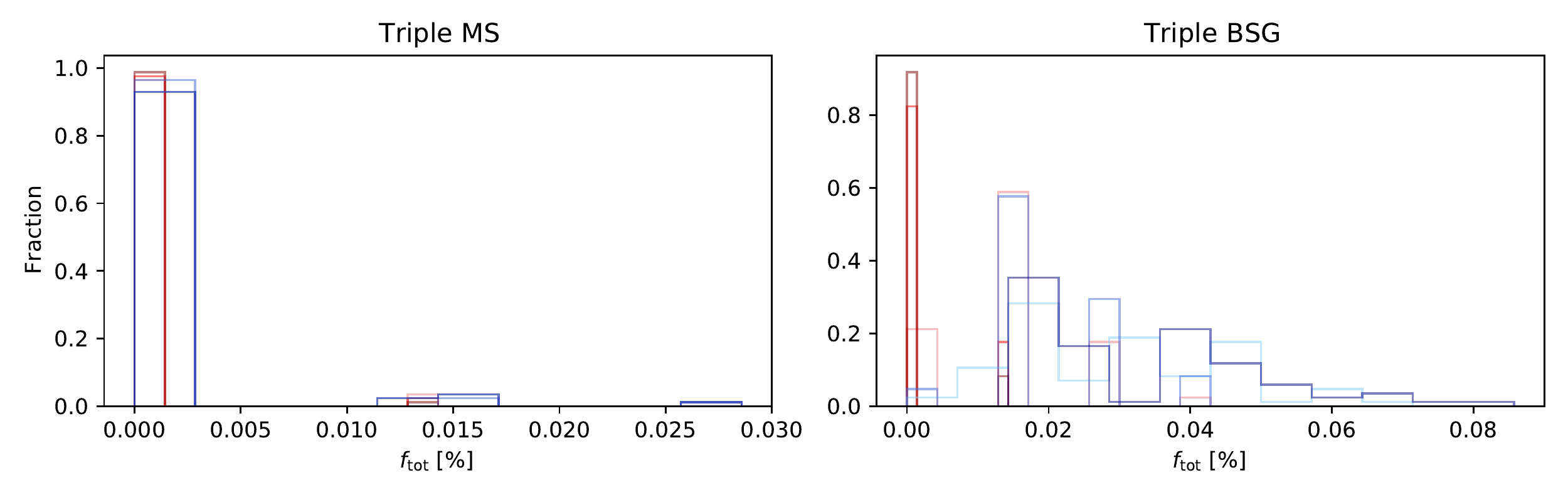}
    \end{subfigure}
    \caption{Total fraction of identified systems when using the MS curves for single-degenerate binaries with a BH (top row), non-degenerate binaries (middle row), and triples from the SMASH+ (bottom row). Left and right panels are for OB and BSG primaries, respectively.}
    \label{fig_mass_uncertainties}
\end{figure*}

\section{The nonnecessity of accurate mass measurements}\label{sec_mass_measurements}
Three of the four parameters necessary to determine the observational AMRF, the middle term of \Eqref{eq_AMRF}, can be obtained from the \Gaias astrometric binary solutions (the astrometric signal, the parallax, and the period). The masses of the primary stars $M_1$ are also provided for a subsample of stars in DR3. However, the DR3 astrophysical parameters of massive stars 
are unreliable due to a lack of quality diagnostic lines in the Gaia spectral range for objects hotter than 10 to 15~kK. Hence, reliable $M_1$ values still need to be determined. There are several ways these masses can be estimated. \nextline
Direct mass measurements can be obtained from eclipsing double-lined spectroscopic binaries. However, these are often short period and most will be too tight to be identified in DR3. For non-eclipsing spectroscopic binaries, only the mass function or minimal mass can be determined instead of the absolute masses \citep[see e.g.][]{Mahy_2020}. Mass estimates can also be obtained from atmospheric analysis (effective temperature and surface gravity) or by comparison to evolutionary tracks. Even the knowledge of spectral type and luminosity class can provide a first estimate by comparison with calibration tables \citep[e.g.][]{Martins_2005}. However, spectroscopic data are not available for the vast majority of the sources in the ALS\,II. \nextline
In principle, effective temperatures and luminosities can also be determined through fitting of a spectral energy distribution. However, the latter is often ambiguous for massive stars as the available photometric data in the visual and UV is in the Rayleigh Jeans domain. \nextline
Since there is no spectrum available at hand for most stars in the ALS\,II, we explored the effect of inaccurate and imprecise mass measurements of the primary on the identification fractions of OB/BSG+BHs and false-positives (i.e. OB/BSG+MS and triple systems). Through a MC simulation, the observational parameters of the simulated systems were varied with the same uncertainties as used in \Secref{sec_population_of_SGs}. We distinguish between several scenarios. Three scenarios assume that the mass of the system is determined, but with different mass errors of 50\%, 100\%, and 200\% (i.e. 0.5 times the mass, one time the mass, and 2 times the mass, respectively). In figures, these scenarios are marked as `0.5Merr', `1Merr', and `2Merr', respectively. The lower and upper limit on the simulated masses was 7 and 60\,$\Modot$, respectively. The other scenarios investigate what would happen to the fractions if we assumed a fixed mass for all systems, for example, fixing all the masses of the primaries to 10$\Modot$, labelled with `10$\Modot$'.\nextline
We discuss the results of the simulations using the MS and BSG curves in Secs. \ref{sec_mass_uncertainty_MScurves} and \ref{sec_mass_uncertainty_BSGcurves}, respectively. While these two sections argue that $M_1$ does not need to be known, the derived mass of the compact object will depend on $M_1$ and might even change the nature of the compact object \citep[e.g. the companion being a 1.2$\Modot$ white dwarf instead of a 1.5$\Modot$ NS; see Sec. 9 in][]{El-Badry_2022_SunBH}. However, for known BH masses, a very large error on the primary mass is needed before the nature of the compact object is changed. For example, in one of the worst-case scenarios, an estimated primary mass of 16$\Modot$ will have a minimum observational AMRF value $\mathcal{A} = 0.37$, corresponding to a minimum $q$ of 0.47. This results in a minimum BH mass of 7.5$\Modot$. If the primary mass is actually 8$\Modot$, the minimum mass of the compact object at the same AMRF value is 5$\Modot$ and hence it is still identified as a BH \citep[see][for the determination of the minumum compact-object mass from $\mathcal{A}$]{Shahaf_2022}.

\begin{figure*}
    \centering
    \begin{subfigure}{\linewidth}
    \includegraphics[width = \textwidth,trim={0 0.95cm 0 0},clip]{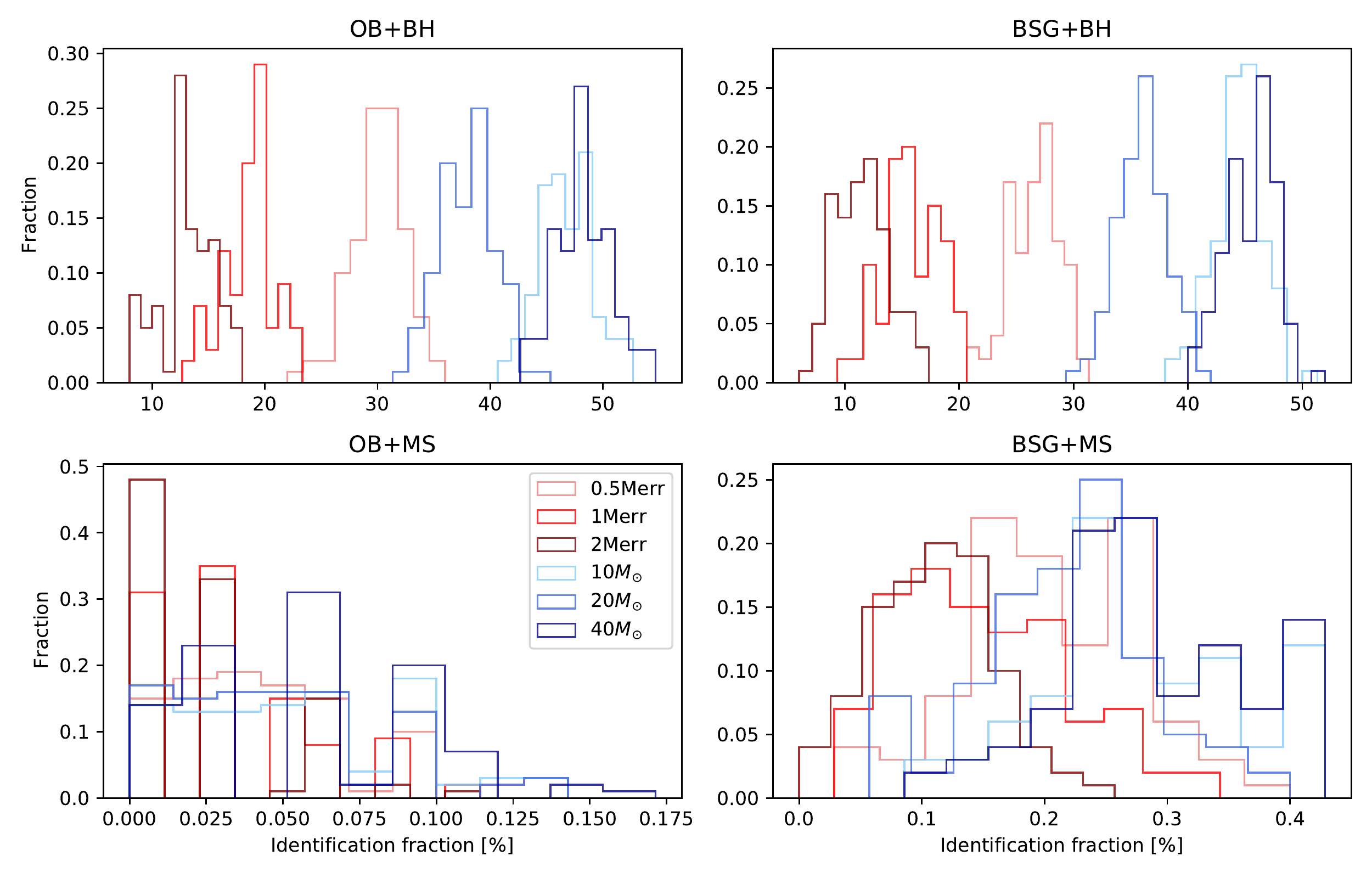}
    \end{subfigure}
    \begin{subfigure}{\linewidth}
    \includegraphics[width = \textwidth]{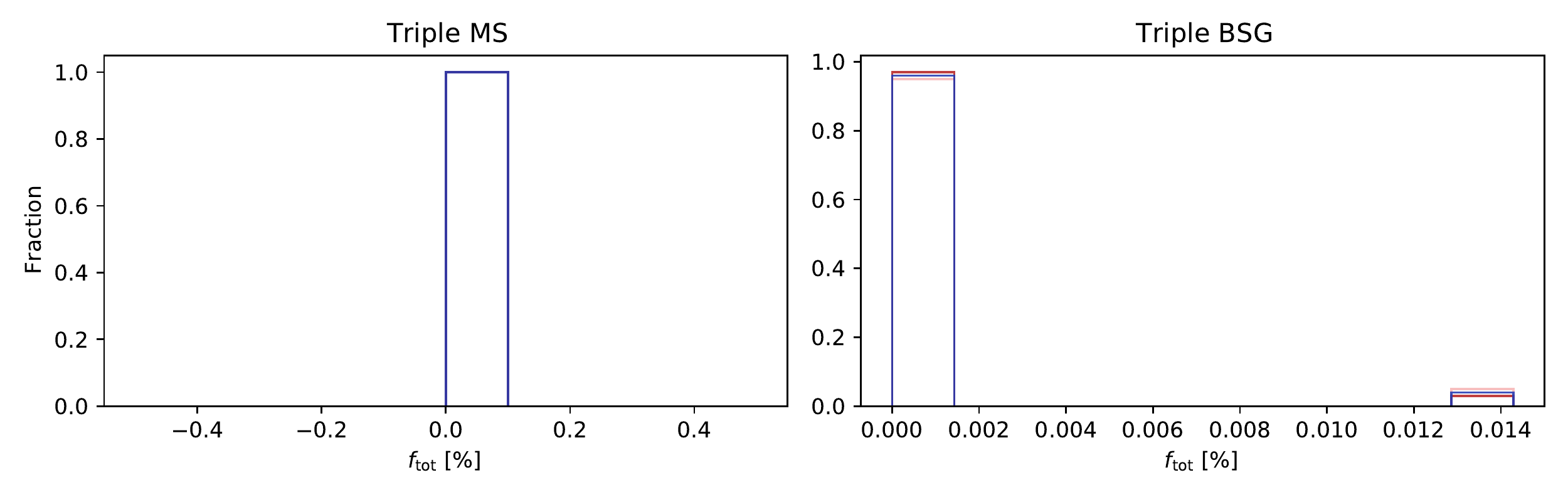}
    \end{subfigure}
    \caption{Same as \Figref{fig_mass_uncertainties}, but when using the BSG curves.}
    \label{fig_mass_uncertainties_SG}
\end{figure*}

\subsection{Total identification fractions using the MS curves}\label{sec_mass_uncertainty_MScurves}
Figure \ref{fig_mass_uncertainties} shows the distribution of total identification fractions (see Eq. \ref{eq_total_fraction}) of 100 samples of 150 OB+BH, 150 BSG+BH, 3500 OB+MS, 3500 BSG+MS, and also two sets of 7000 triples from the SMASH+ with OB and BSG primaries using the MS curves. The total identification fractions show a trend for single-degenerate systems with a BH (OB/BSG+BH systems, top row of \Figref{fig_mass_uncertainties}). For the three scenarios assuming a determined mass with an error (`0.5Merr', `1Merr', and `2Merr') there is a clear difference. The larger the error on the primary mass, the larger the error on $\mathcal{A}$, hence the smaller the identification fraction (the identification criterion of \citetalias{Janssens_2022} was $\mathcal{A}-\sigma_{\mathcal{A}} > \mathcal{A}_{\text{triple}}$). The total identification fractions are even slightly higher when using a fixed mass for all primaries, since no error on the mass is assumed in this case.\nextline
The value of the fixed mass does not significantly affect the total identification fractions of the single-degenerate systems. When assuming a different primary mass, the curves in the AMRF diagram will change, going downwards for higher primary masses (see \Figref{fig_AMRF_curves_SJ22}). A higher primary mass also decreases $\mathcal{A}$ (see \Eqref{eq_AMRF}). When altering the primary mass, the rate with which both the MS curves and $\mathcal{A}$ move downward in the AMRF diagram is comparable over most of the massive-star dwarf mass range that was derived here ($8<M/\Modot\lesssim 60$). Hence, the value of the fixed mass does not play a large role for the identification of OB/BSG+BHs.\nextline
The above described effect of larger uncertainties on the mass or a fixed mass is almost not present for the false-positive identification of OB+MS binaries (middle-left panel of \Figref{fig_mass_uncertainties}). The reason for this is that OB+MS systems are located very low in the AMRF diagram. Therefore, it is already very difficult for these systems to be falsely identified as a single-degenerate binary. Hence, the effect of larger mass uncertainties or zero mass uncertainties becomes barely prominent.\nextline
For the false-positive BSG+MS binaries, we see that a larger uncertainty on the mass does have the same effect as for the single-degenerate binaries. However, we now also see a change in fractions for different values of the fixed mass. The mass-magnitude relation for BSGs (see \Figref{fig_mass_magnitude}) is rather steep for $8<M/\Modot<30$ and flattens for $M\gtrsim 30\Modot$, but the mass-magnitude relation for MS stars (see the gray stars in \Figref{fig_mass_magnitude} and figure 3 in \citetalias{Janssens_2022}) does not. Over almost the whole massive mass regime, the mass-magnitude relation for MS stars retains almost the same slope, except for $8<M/\Modot<15$, where the slope changes more significantly. Because of the combination of the two mass-magnitude relations, BSG+MS binaries with different primary masses take up significantly different positions in the AMRF diagram compared to their OB+MS counterparts. The effect of this is non-trivial and is most likely the cause of the change in total identification fractions for different values of the fixed mass. However, each value of the fixed mass does still result in total identification fractions with the same order of magnitude.\nextline
Finally, we can see that a fixed value for the mass slightly increases the total identification fraction on average for SMASH+ triples with a BSG primary because of the same effect as with the BSG+MS binaries. Overall, the total identification fraction of the SMASH+ triples is extremely low, both for OB primaries as well as BSG primaries, as expected from \Secref{sec_triple_population}.\nextline
When using the MS curves, an accurate mass estimate is not necessary to identify OB/BSG+BH binaries. In fact, a fixed value of the mass will lead to slightly higher total identification fractions because in that case no error on the mass is assumed. The order of magnitude of the false-positive identification among the BSG+MS binaries remains the same and below 5\% for any explored scenario, but can vary depending on the value of the fixed mass.

\subsection{Total identification fractions using the BSG curves}\label{sec_mass_uncertainty_BSGcurves}
The total identification fractions for OB/BSG+BH binaries become much lower when using the BSG curves (\Figref{fig_mass_uncertainties_SG}, top panels), as already shown in \Secref{sec_population_of_SGs}. As for \Secref{sec_mass_uncertainty_MScurves}, a larger uncertainty on the mass results in lower fractions. This time however, we can also see here that different values for the fixed mass somewhat change the fractions, even for the OB+MS systems. The decrease of the maximum of the theoretical BSG curves and $\mathcal{A}$ is no longer comparable and hence causes a change in the fractions when using different values for the fixed mass. \nextline
The total false-positive identification fractions of BSG+MS binaries using the BSG curves (\Figref{fig_mass_uncertainties_SG}, middle-right panel) now behaves the same as those for OB+MS curves when using the MS curves (\Figref{fig_mass_uncertainties_SG}, middle-left panel). They are extremely small, reaching maximum values of $\sim0.4$\%. The total identification fractions for OB+MS binaries are even lower when using BSG curves, with a maximum below 0.2\%.\nextline
The triples in the SMASH+ are not resulting in any false-positives (\Figref{fig_mass_uncertainties_SG}, bottom panels). Sometimes however, one system ($\sim0.014$\%) is falsely identified when using the BSG curves.\nextline
When using the BSG curves, the choice of the value of a fixed mass can slightly affect the total identification fractions. However, in all panels, we can see that the order of magnitude of the fractions remains the same. Especially for the false-positive identifications, the most important result is that the value of the fixed mass will not significantly increase nor decrease the fraction of false-positives.

\section{Results from \Gaias DR3}\label{sec_gaia_DR3_results}
On June 13th 2022, the third \Gaias data release DR3 was published \citep[][]{DR3_2022}, including the first \Gaias astrometric binary solutions \citep[][]{Halbwachs_2022}. In total, the DR3 catalogue contains astrometric binary solutions for 169\,227 sources. \nextline
The assumptions adopted in \citetalias{Janssens_2022} led to an estimated 2000 sources in the ALS\,II having an astrometric binary solution in DR3 (see \Tabref{table_numbers}). Performing a cross-match between the ALS\,II and the DR3 astrometric binary sources table resulted in 11 sources, none of which satisfy our identification criterion of $\mathcal{A}-\sigma_{\mathcal{A}} > \mathcal{A}_{\text{triple}}$. \nextline
Although we find no strong OB+BH candidates, the non-detection reported here cannot be interpreted as a result of the BH-formation scenario, even though for some BH-formation scenarios we expected barely any detections. Instead, the small number of massive binaries with an astrometric binary solution in DR3 is in line with new predictions using the selection criterion imposed by the \Gaias collaboration, which we elaborate on in \Secref{sec_parallax_precision}. We investigate how much less conservative the selection criteria need to be in order to obtain information on the BH-formation scenario in \Secref{sec_less_conservative}. In \Secref{sec_low_mass_DR3}, we discuss briefly the work done for low-mass single-degenerate systems.

\subsection{The impact of the DR3 selection criteria on the number of published astrometric binaries}
\label{sec_parallax_precision}
There are several selection criteria for publishing astrometric binary solutions listed in \citet{Halbwachs_2022}. Here, we investigate the impact of two of those.
\subsubsection*{The required precision on $\boldsymbol{\varpi}$}
\begin{figure}
    \centering
    \includegraphics[width = \linewidth]{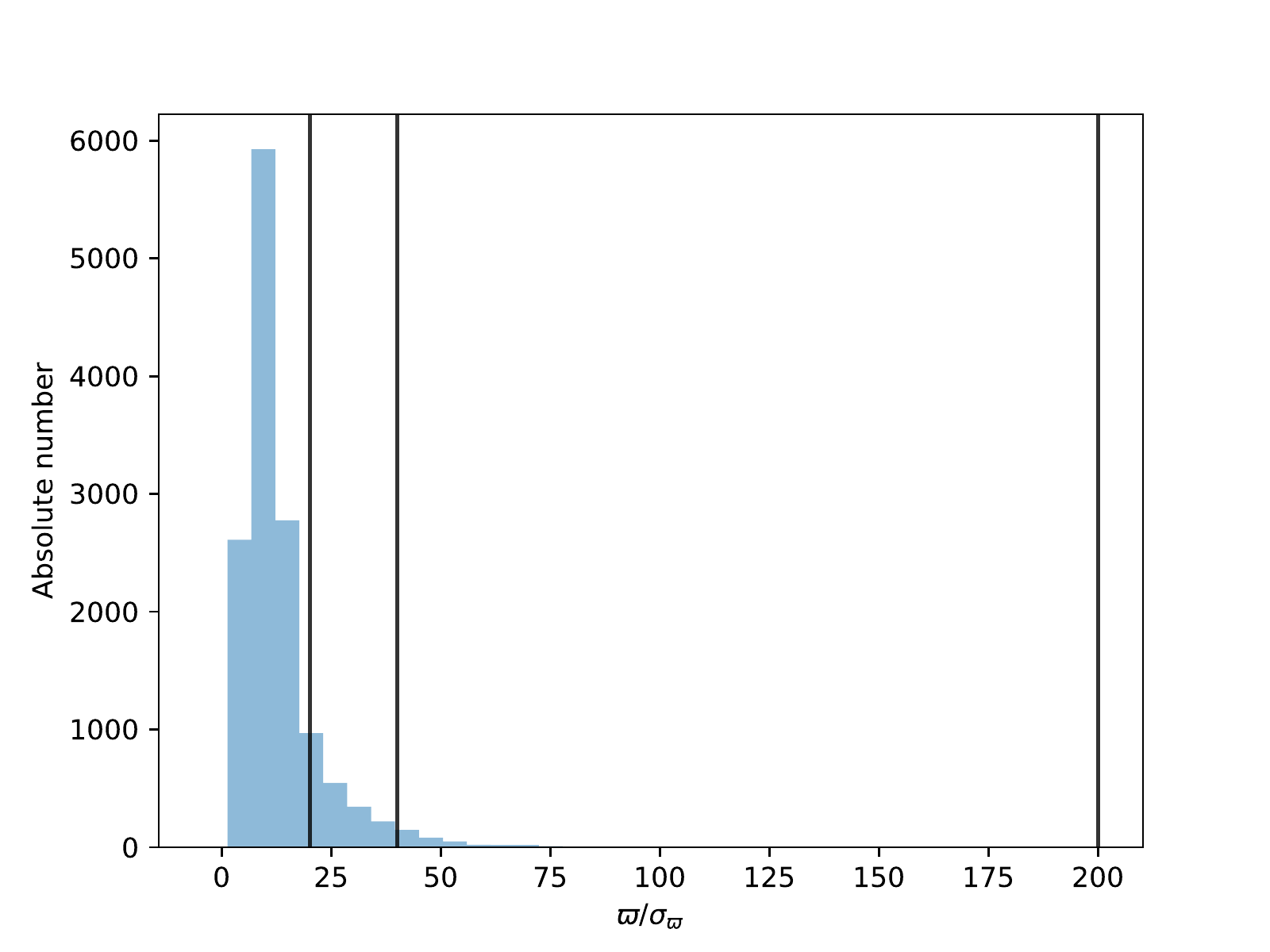}
    \caption{Histogram of $\varpi/\sigma_{\varpi}$ of the massive sources in the ALS\,II. The vertical lines indicate a precision on $\varpi$ of 200, 40, and 20.}
    \label{fig_parallax_over_error_alsii}
\end{figure}

One of the criteria for an astrometric binary solution to be published in DR3 is related to the relative parallax precision:
\begin{equation}\label{eq_Gaia_criterion}
    \varpi/\sigma_{\varpi} > 20\,000/P_{\text{day}},
\end{equation}
where $\varpi$ and $\sigma_{\varpi}$ are the parallax and its error and $P_{\text{day}}$ is the derived period in days. For orbital periods of 100, 500, and 1000 days, this criterion implies a relative precision on $\varpi$ of 200, 40, and 20, respectively. The majority of the OB+BH systems are expected to have $P = 100-500$\,days.
\nextline
In order to investigate how many OB+BH sources we would expect in the DR3 astrometric binary catalogue assuming only the criterion in \Eqref{eq_Gaia_criterion}, we performed the same simulations presented in \citetalias{Janssens_2022} now with a different detectability criterion. We obtained parallaxes and parallax uncertainties from the \Gaias catalogue for all sources in the ALS\,II. Instead of only drawing distances for the simulated OB+BH systems, we now randomly draw for each system a parallax and its uncertainty, such that we can determine $\varpi/\sigma_{\varpi}$ (this also equals the parameter \texttt{parallax\_over\_error} in the \Gaias catalogue). The distribution of $\varpi/\sigma_{\varpi}$ of the sources in the ALS\,II is given in \Figref{fig_parallax_over_error_alsii}. There are no systems with $\varpi/\sigma_{\varpi}> 200$.

Using \Eqref{eq_Gaia_criterion} as the detectability criterion for the simulated OB+BH systems, we find that only 0.14\% are detected. Assuming 300 OB+BH systems in the ALS\,II, this would result in 0 or 1 detection. Similarly, for non-degenerate binaries, only 0.3\% are detectable. If there are 7000 binaries in the ALS\,II, this leads to the detection of $\sim$20 astrometric binaries. This predicted amount is very close to the 11 ALS\,II sources with astrometric binary solutions in DR3. Hence, the non-detection of OB+BH candidates is expected from the adopted cut and does not teach us anything yet on the underlying BH-formation scenario.

\subsubsection*{The required precision on the derived astrometric signal}
Another criterion is limiting the precision on the derived astrometric signal $\alpha_0$ or the significance $s$:
\begin{equation}\label{eq_Gaia_criterion_significance}
    s = \alpha_0/\sigma_{\alpha_0} > 158/\sqrt{P_{\text{day}}},
\end{equation}
where $\sigma_{\alpha_0}$ is the uncertainty on $\alpha_0$. When using a significance of 10 (i.e. assuming a 10\% uncertainty on $\alpha_0$), our simulations show that, after applying the criterion in \Eqref{eq_Gaia_criterion}, no additional non-degenerate massive binaries are filtered out. Hence, the detected non-degenerate massive-binary fraction in our simulations remains 0.3\% or $\sim$20 astrometric binaries. On average, the significance of the astrometric DR3 binaries in the ALS\,II is on average $\sim$3 (i.e. a 3\% uncertainty on $\alpha_0$). Hence, in our simulations, the criterion in \Eqref{eq_Gaia_criterion_significance} did not affect the number of published astrometric massive binaries in DR3.\nextline

\subsection{Towards less conservative constraints}\label{sec_less_conservative}
As shown in \Secref{sec_parallax_precision}, the criterion in \Eqref{eq_Gaia_criterion} is the most stringent one. This criterion has been imposed on the DR3 data to avoid many spurious solutions, which otherwise could result in the false-positive detection of many massive BH binaries. While the restriction also filtered out a lot of probably non-spurious solutions \citep[see Figs. 3a and b in][]{Halbwachs_2022}, a significant amount of false-positive sources could have lead to wrong conclusions and many false reports. If this issue is resolved in DR4, \Eqref{eq_Gaia_criterion} can be relaxed, opening up the possibility to discover a large population of BHs. Below we elaborate on how much less conservative the constraints need to be to obtain information on the BH-formation scenario.\nextline
We performed simulations with several constraints on the parallax precision while keeping the constraint on the significance as presented in \Eqref{eq_Gaia_criterion_significance}. Here, $\varpi$ and $\sigma_{\varpi}$ are randomly drawn from the ALS\,II catalogue and the relative uncertainty on $\alpha_0$ is assumed to be 3\%.\nextline
As expected, we find that the constraints need to be significantly lowered to detect a significant amount of BHs. In order to be able to detect $\sim$20 OB+BH binaries or $\sim$7\% of the simulated population, the constraint in \Eqref{eq_Gaia_criterion} needs to be decreased by 75\% (i.e. requiring that $\varpi/\sigma_{\varpi} > 5000/P_{\text{day}}$). Here, it is assumed BHs form with no natal kick and no mass loss. However, there are several scenarios which lead to less OB+BH binaries when, for example, the BH receives a strong kick (see \citetalias{Janssens_2022} for the impact of different BH-formation scenarios on the predicted amount of OB+BHs).\nextline
In order to detect a sensible amount of OB+BH systems in other extreme scenarios involving strong kicks, we would argue for even lower constraints. 
We suggest using $\varpi/\sigma_{\varpi} > 1000/P_{\mathrm{day}}$ or $\varpi/\sigma_{\varpi} > 5$, resulting in 3 identifications in the ``worst-case'' scenario (strong kick), while $>$30 identifications in other scenarios.\nextline
When lowering the required relative precision on $\varpi$, and keeping the relative precision on $\alpha_0$ assumed to be 3\%, the significance criterion (\Eqref{eq_Gaia_criterion_significance}) still does not impact the number of detectable OB+BHs. This shows that the precision on the parallax remains the most critical cut and the significance criterion does not severely impact the number of detectable OB+BHs.\nextline
We did however not take into account that the relative precision on $\varpi$ improves in the astrometric binary solution and might improve still in DR4 when more data is available. In that case, our imposed constraints can easily be scaled.\nextline
To conclude, our simulations showed that the constraint on the relative parallax precision needs to be significantly lowered. If the relative parallax precision of the DR3 single star solution catalogue is used, a decrease by 95\% is needed. Otherwise, if the relative precision improves due to more data or the binary solution, the maximum conservativeness the constraint can have is 
\begin{equation}
  n \times (\varpi/\sigma_{\varpi})_{\text{DR3,single}} > n\times 1000/P_{\text{day}},
\end{equation}
with $n$ the improvement on the relative parallax precision compared to that of the DR3 single star solutions. The constraint on the significance does not necessarily have to be lowered.

\subsection{Lower-mass single-degenerate binaries with Gaia DR3}\label{sec_low_mass_DR3}
While the manuscript focuses on massive single-degenerate binaries with a BH component, it is worth mentioning that some candidates of low-mass single-degenerate binaries with a BH component (i.e. BH systems that have a luminous companion with an initial mass $<8\MSun$) are reported in several other manuscripts. Already in \citet{Arenou_2022}, one of the \Gaias verification papers, the authors use the AMRF on a sample of DR3 astrometric binaries with primary masses derived by \Gaias in the range of $0.2-2\MSun$. While most of their candidates are likely white dwarfs, they also report on a neutron star candidate. However, \citet{El-Badry_2022_SunBH} argue that this candidate is also a white dwarf due to an ambiguity in the mass of the primary star.\nextline
Currently, there has been one report on a candidate BH system orbiting around a solar-type star \citep[][]{El-Badry_2022_SunBH}, which was almost filtered out by \Eqref{eq_Gaia_criterion}. While the solution showed signs of potentially being spurious, follow-up spectroscopic data showed that the astrometric orbit derived by \Gaias is indeed correct. This candidate was also picked up by \citet{Shahaf_2022} as a candidate system containing a BH with the AMRF method. \citet{Shahaf_2022} also report on some NS and white dwarf candidates.\nextline
Different research teams also used the other available \Gaias binary solutions (i.e. photometric and spectroscopic), in the search for compact objects. While some photometrically detected single-degenerate binary candidates are reported \citep[][]{Gomel_2022}, several studies conclude there are no spectroscopic candidates \citep[][]{El-Badry_2022_SB1s, Fu_2022, Jayasinghe_2022}.

\section{Summary}\label{sec_summary}
We have investigated how different biases and uncertainties may affect the number of expected OB+BH identifications predicted in \citetalias{Janssens_2022}. We have expanded their work by including the effect of a different mass-magnitude relation on the total identification fractions (i.e. the fraction of systems detectable by \Gaias and identifiable with the AMRF). For this, we have derived a mass-magnitude relation for BSG stars, resulting in the BSG curves.\nextline
We found that the contribution of OB+MS binaries to the amount of false-positives is negligible and around 0.1\% when using the mass-magnitude relation for OBs derived in \citetalias{Janssens_2022}, resulting in the MS curves. This fraction decreases when using the BSG curves. With an estimated 3500 of OB+MS binaries in the ALS\,II, the number of false-positive systems would be between $\sim$2-6 depending on the used curves. The fraction of false-positive BSG+MS systems is $\sim$1.3\% when using the MS curves and drops to less than 1\% when using the BSG curves. For an estimated 3500 BSG+MS binaries in the ALS\,II, this would lead to $\sim$50 or $\sim$2 false-positive identifications. Comparing this to the expected $\sim$200 or $\sim$85 OB/BSG+BH identifications using MS or BSG curves, we can see that when including BSG primaries, it is recommended to use the BSG curves to increase the fraction of true identifications to false-positives.\nextline
We further widened the contribution of false-positives to triple systems and higher-order multiple systems. First, we simulated a grid of triples where the outer star is more luminous than the inner binary. There is a large difference between using the MS curves or the BSG curves especially for systems with a BSG primary. When using the MS curves, the false-positive fraction in some parts of the explored grid can go up to 50\%. Although these kind of triples are not expected to be very abundant, 50\% is a highly non-negligible false-positive fraction. Instead, when using the BSG curves, the maximum false-positive fraction drops to around $\sim$2.5\%. Here, the usage of the BSG curves can thus highly reduce the amount of false-positives.\nextline
We also simulated a mock population using the triples in the SMASH+ both for OB and BSG primaries. Most of the triples in the SMASH+ have inner binaries which are more luminous than the outer star. Hence, the false-positive fraction of triples in the SMASH+ is extremely low and maximum around 0.01\% both when using the MS or BSG curves, potentially leading to one false-positive identification. This is a highly negligible amount.\nextline
The contribution of quadruple or higher-order multiple systems should be extremely low. This is mostly due to the chaotic motion of the photocentre and the generally long outer periods ($P>5$yr).\nextline
Finally, we have shown that an accurate mass measurement is not very important for the identification of OB/BSG+BH binaries, nor for the amount of false-positives. We have experimented both with different errors on the mass as well as fixing the mass of the primary to the same value for the whole population. As expected, larger errors result in lower identification fractions. A fixed mass however results in even slightly higher identification fractions for the OB/BSG+BH binaries and, most importantly, not in significantly different false-positive fractions. This result is not any different when using the MS curves or the BSG curves.\nextline
Ideally, the evolutionary stage and mass of the primary star are known. This way, the corresponding mass-magnitude relation can be used to establish the AMRF curve corresponding to the correct primary mass. If the evolutionary stage is not known, which is the case for most of the stars in the ALS\,II, then it can be recommended to use the BSG curves instead of the MS curves, especially to lower the false-positive identification fractions of triples where the outer star is a luminous BSG.\nextline
Of the order of 10 ALS\,II sources have an astrometric binary solution in DR3, of which none are good OB+BH candidates. Although this null-result was predicted for some BH-formation scenarios, the low number of ALS\,II sources with an astrometric binary solution is likely caused by the stringent selection criterion imposed on $\varpi/\sigma_{\varpi}$ by the \Gaias collaboration.\nextline
Few extra compact-object candidates are found among the large pool of data given by DR3. It is clear that the search for single-degenerate binaries in any mass regime remains a hot topic among several research groups. While no detection criteria are known yet for DR4, the relaxation of the \Gaias detection criteria could tremendously improve the \Gaias discovery capability for single-degenerate binaries, especially among the massive stars. Our simulations showed that the constraint on the relative parallax precision needs to be lowered by 95\% to obtain information on the BH-formation scenario. We therefore suggest that, the most conservative the constraint can be in DR4 is $n \times (\varpi/\sigma_{\varpi})_{\text{DR3,single}} > n\times 1000/P_{\text{day}}$, with $(\varpi/\sigma_{\varpi})_{\text{DR3,single}}$ the relative parallax precision for the single source solution in DR3 and $n$ the DR4 improvement of the relative precision compared to the DR3 single star solutions.

\begin{acknowledgements}
SJ acknowledges support from the FWO PhD fellowship under project 11E1721N. 

This project has received funding from the European Research Council (ERC) under the European Union's Horizon 2020 research and innovation programme (grant agreement n$\degree$ 772225/MULTIPLES). TS acknowledges support from the European Union's Horizon 2020 under the Marie Skłodowska-Curie grant agreement No 101024605. 

This work has made use of data from the European Space Agency (ESA) mission
{\it Gaia} (\url{https://www.cosmos.esa.int/gaia}), processed by the {\it Gaia}
Data Processing and Analysis Consortium (DPAC,
\url{https://www.cosmos.esa.int/web/gaia/dpac/consortium}). Funding for the DPAC
has been provided by national institutions, in particular the institutions
participating in the {\it Gaia} Multilateral Agreement.
\end{acknowledgements}

\bibliography{References}

\newpage
\begin{appendices}
\renewcommand{\thefigure}{A\arabic{figure}}
\setcounter{figure}{0}
\renewcommand{\thetable}{A\arabic{table}}
\setcounter{table}{0}

\section{Masses, effective temperatures, and radii}\label{sec_appendix_MTR}
Table \ref{table_data_mass-mag} shows the literature data obtained for BSGs.
\begin{table*}[!h]
    \centering
    \caption{Masses, effective temperatures and radii of BSGs used for the fitting of the mass-magnitude relation in \Secref{section_magnitudes_evolved}. If available, spectral types or initial masses and rotational velocities are also given.}
     \begin{tabular}{lrclrcl}
        \hline \hline
        $M$ [$\Modot$] & $\Teff$ [K]& $R$ [$\Rodot$]& SpT & $M_{i}$ [$\Modot$]& $v_{\text{rot},i}$ [km s$^{-1}$]&Ref.\\
        \hline
       66.89 & 42551 & 18.47 & O3\,I &&& a\\
       58.03 & 40702 & 18.91 & O4\,I &&& a\\
       50.87 & 38520 & 19.48 & O5\,I &&& a\\
       48.29 & 37070 & 19.92 & O5.5\,I &&& a\\
       45.78 & 35747 & 20.33 & O6\,I &&& a\\
       43.10 & 34654 & 20.68 & O6.5\,I &&& a\\
       40.91 & 33326 & 21.14 & O7\,I &&& a\\
       39.17 & 31913 & 21.69 & O7.5\,I &&& a\\
       36.77 & 31009 & 22.03 & O8\,I &&& a\\
       33.90 & 30504 & 22.2 & O8.5\,I &&& a\\
       31.95 & 29569 & 22.6 & O9\,I &&& a\\
       30.41 & 28430 & 23.11 & O9.5\,I &&& a\\
       28.19 & 26903 & 22.29 &  & 30.0 & 269& b\\
       23.95 & 25589 & 20.67 & &25.0&271& b\\
       19.52 & 24111 & 18.03 &  &20.0 & 274 & b\\
       14.80 & 22155 & 14.70 &  &15.0&277& b\\
       11.87 & 20475 & 12.38 &  &12.0&279& b\\
       9.89 & 18818 & 11.03 &  &10.0&281& b\\
       8.91 & 7843 & 10.33 &  &9.0&281& b\\
        \hline
     \end{tabular}
         \flushleft
    \begin{tablenotes}
      \small
      \item \textbf{Notes.} Data from: $^{\text{(a)}}$ \cite{Martins_2005}; $^{\text{(b)}}$ \citet{Brott_2011}
    \end{tablenotes}
    \label{table_data_mass-mag}
\end{table*}

\newpage

\section{Using a 10\% vs 1\% core-H fraction for the BSGs}\label{appendix_10_vs_1}
\begin{figure}[!b]
    \centering
    \includegraphics[width = \linewidth]{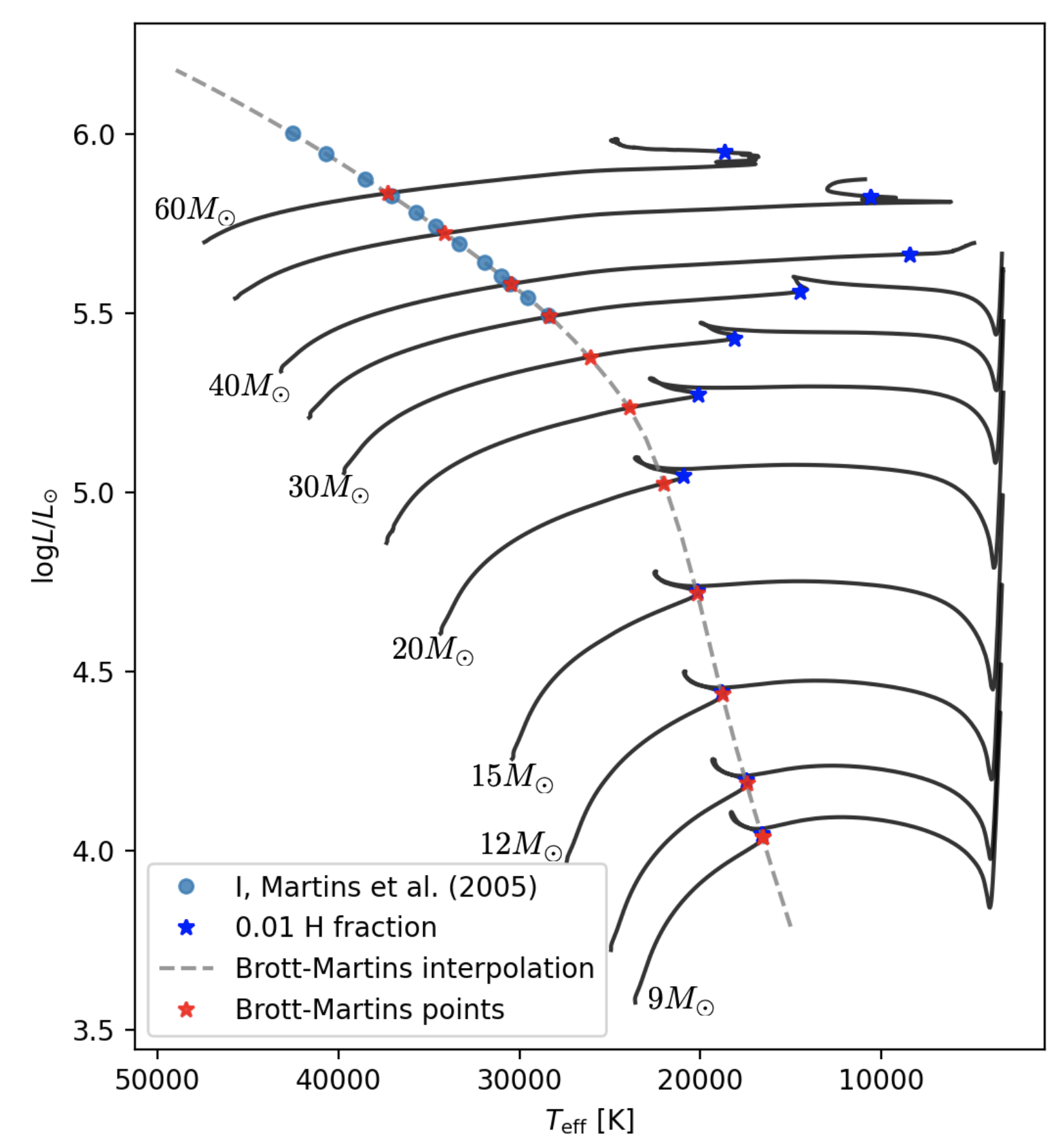}
    \caption{Same as \Figref{fig_magnitudes_evolved} but using a 1\% core-H fraction.}
    \label{fig_magnitudes_evolved_1}
\end{figure}

\begin{figure*}
    \centering
    \includegraphics[width = \linewidth]{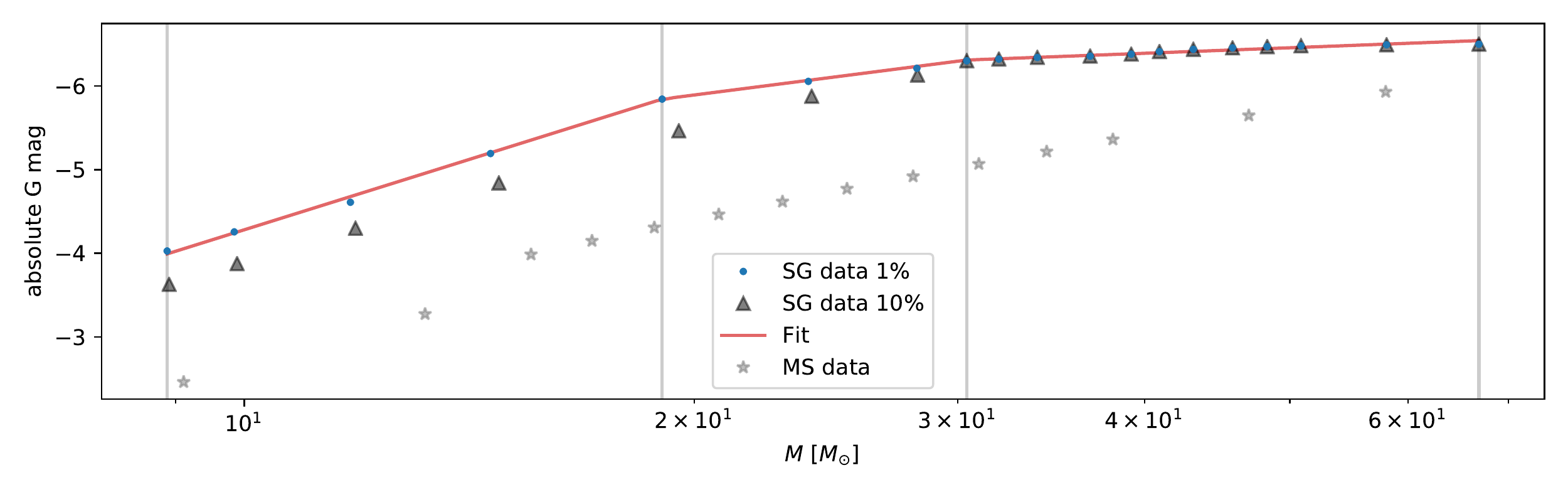}
    \caption{Same as \Figref{fig_mass_magnitude} but using a 1\% core-H fraction. The absolute $G$ magnitudes using a 10\% core-H fraction are also shown as a reference.}
    \label{fig_mass_magnitude_1}
\end{figure*}

\begin{figure*}
    \centering
    \begin{subfigure}{0.5\linewidth}
    \includegraphics[width = \textwidth]{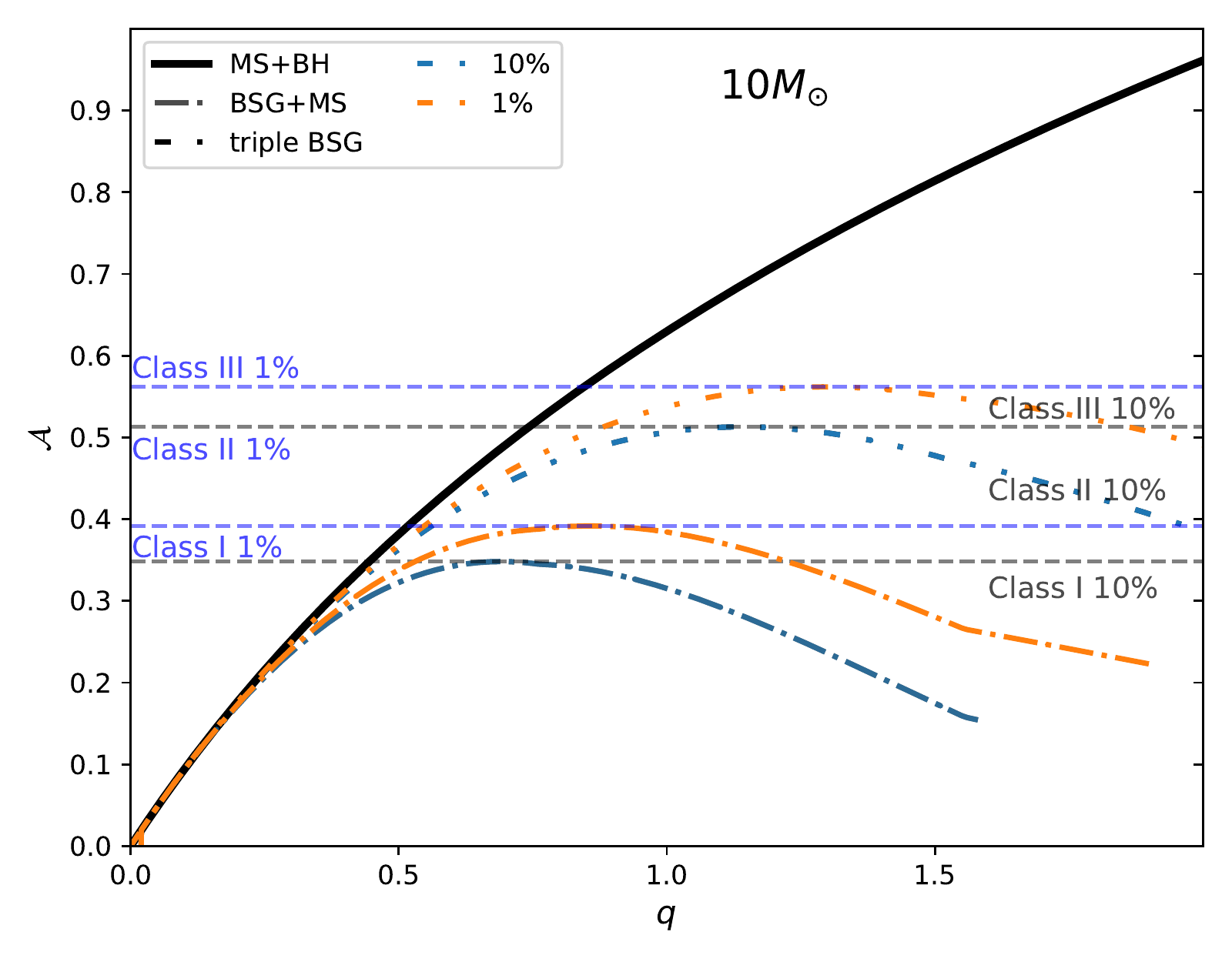}
    \end{subfigure}
    \hspace{-0.1in}
    \begin{subfigure}{0.5\linewidth}
    \includegraphics[width = \textwidth]{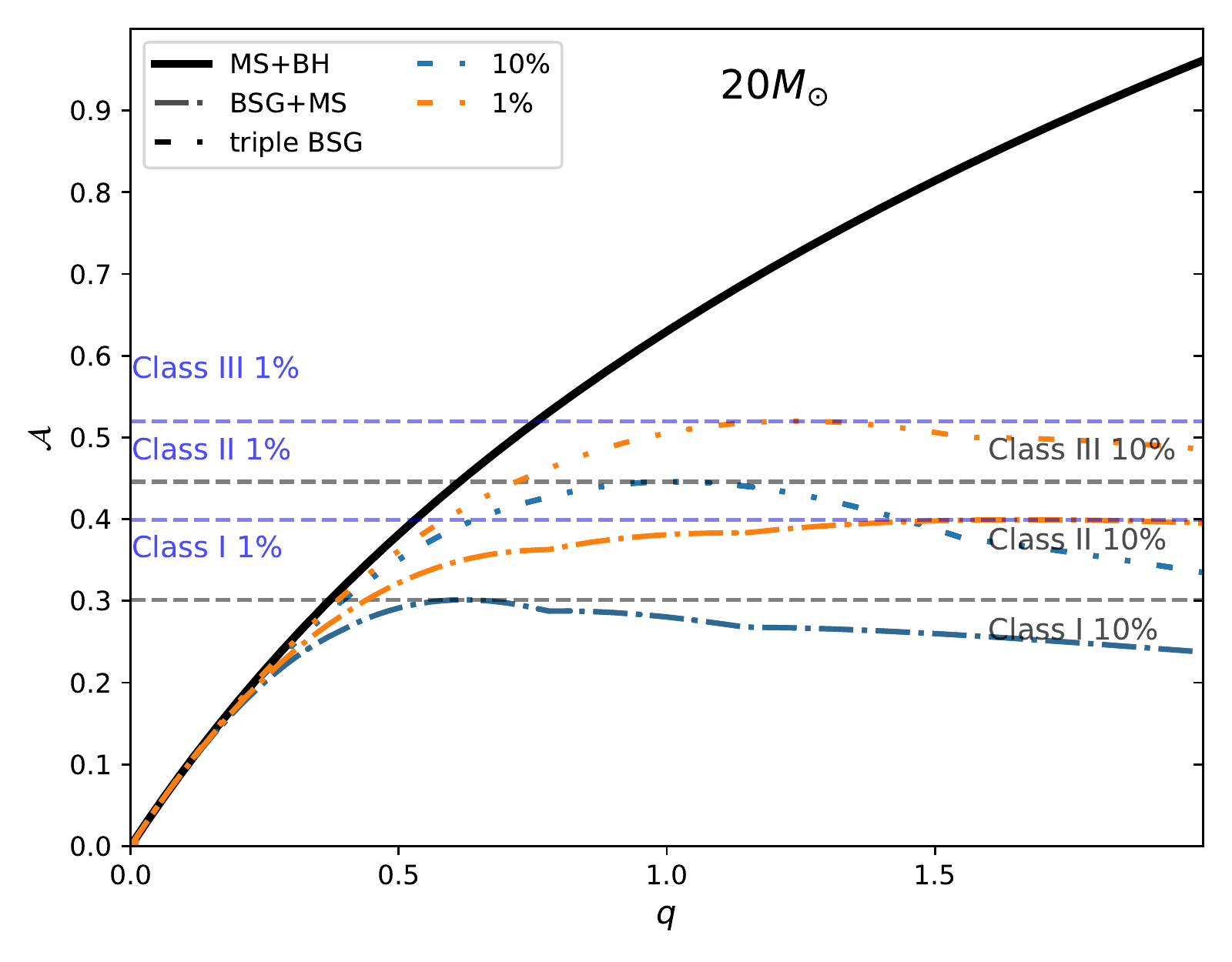}
    \end{subfigure}
    \caption{Theoretical AMRF curves for a 10\,$\Modot$ (left) and 10\,$\Modot$ (right) primary BSG star obtained from the mass-magnitude relation using a 10\% (blue) or 1\% (orange) core-H fraction. The different classes in both cases are also marked.}
    \label{fig_AMRF_curves_SGs_compare}
\end{figure*}

Here, we show the mass-magnitude relation for BSG stars derived as described in \Secref{section_magnitudes_evolved} using a 1\% core-H fraction instead of 10\% (Figs. \ref{fig_magnitudes_evolved_1} and \ref{fig_mass_magnitude_1}). The fit parameters are listed in \Tabref{table_G-mag_fit_1}. We also show an example of the difference in theoretical AMRF curves between these two cases (\Figref{fig_AMRF_curves_SGs_compare}).

\begin{table}
\centering
\caption{Fit parameters for the mass-magnitude relation of BSGs in different mass regimes using a 1\% core-H fraction.}
\begin{tabu}{ ccc }
     \hline
     \hline
     $M_{\text{low}}-M_{\text{up}}$ [$\Modot$] & $a$ & $b$\\ 
     \hline
     30.41 - 66.89 & $-0.72 \pm 0.01$ & $-5.23 \pm 0.02$\\ 
     19.03 - 30.41 & $-2.29 \pm 0.01$ & $-2.91 \pm 0.02$\\ 
     8.9 - 19.03 & $-5.59 \pm 0.01$ & $1.31 \pm 0.01$\\
     \hline
     \end{tabu}
\label{table_G-mag_fit_1}
\end{table}

\newpage~\newpage

\section{Identification fractions for a grid of triples}
\begin{figure*}
    \centering
    \begin{subfigure}{\linewidth}
    \includegraphics[width = \textwidth]{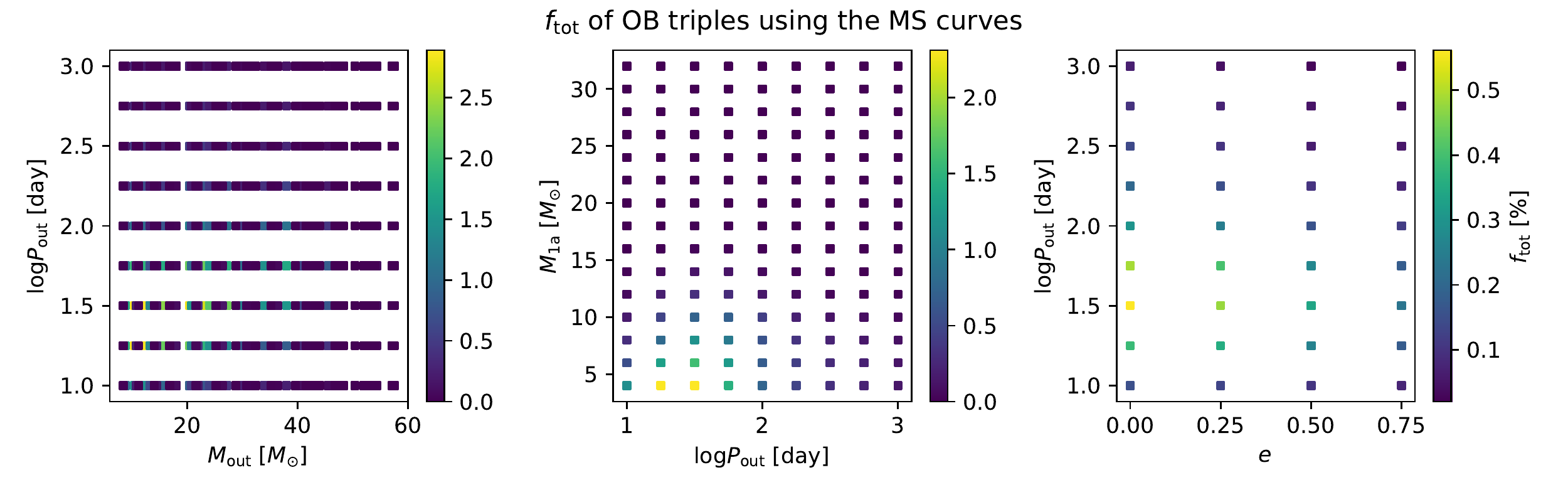}
    \end{subfigure}
    \hspace{-0.1in}
    \begin{subfigure}{\linewidth}
    \includegraphics[width = \textwidth]{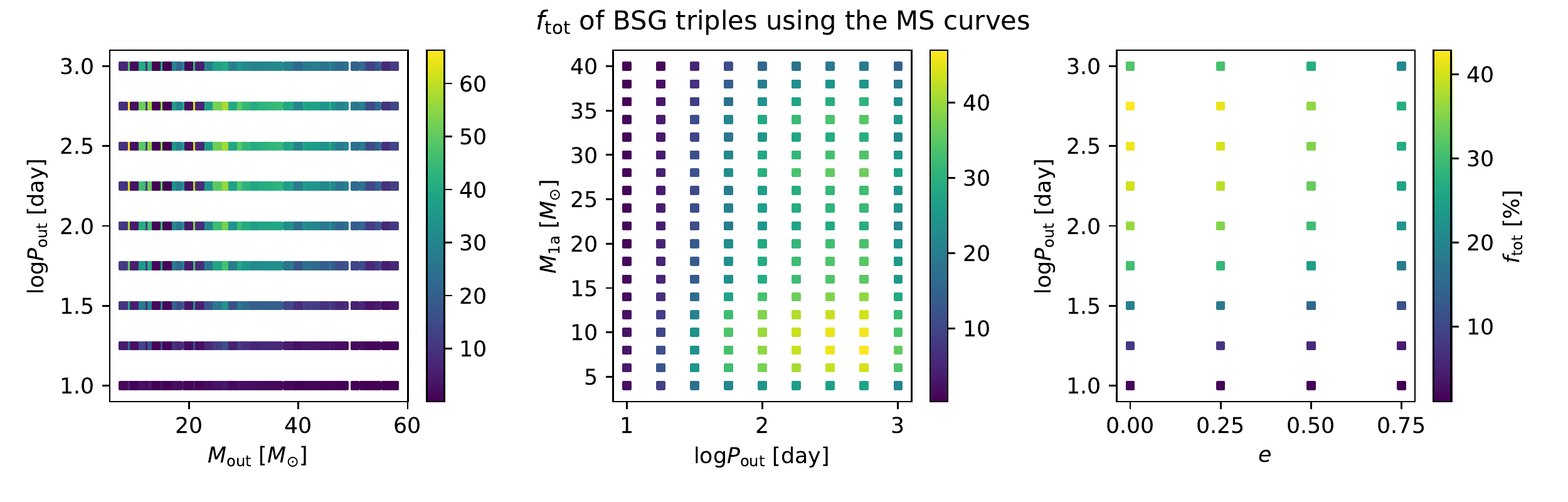}
    \end{subfigure}
    \caption{Same as \Figref{fig_grid_identified_part_OB}, but for more parameters.}
    \label{fig_grid_OBs_full}
\end{figure*}

\begin{figure*}
    \centering
    \begin{subfigure}{\linewidth}
    \includegraphics[width = \textwidth]{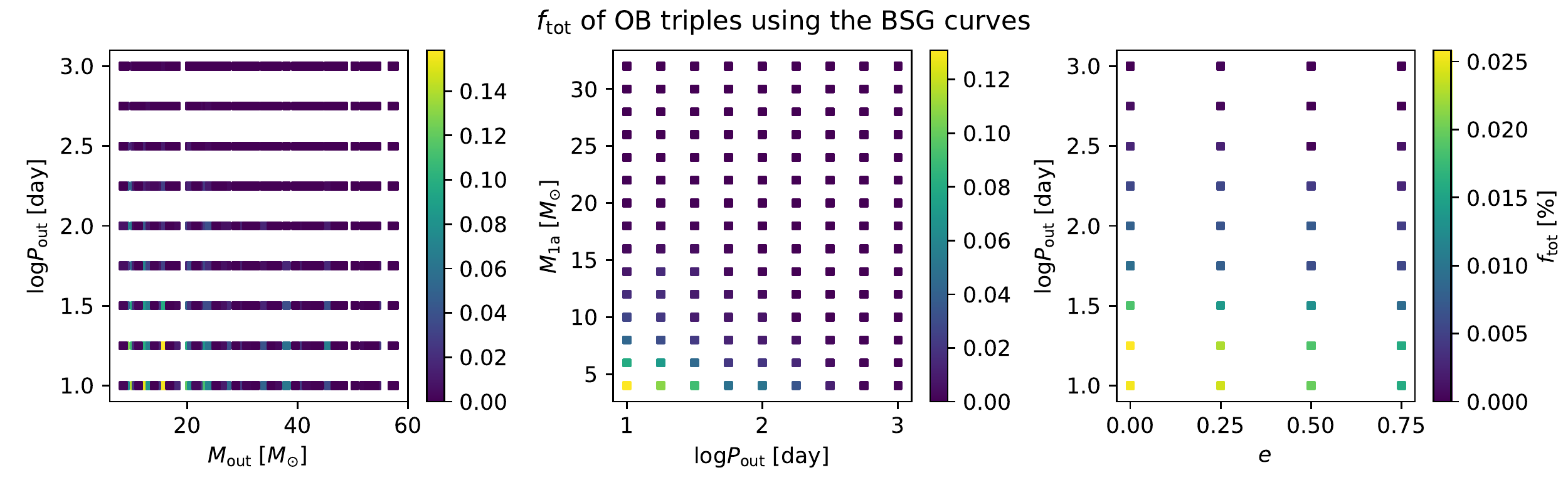}
    \end{subfigure}
    \hspace{-0.1in}
    \begin{subfigure}{\linewidth}
    \includegraphics[width = \textwidth]{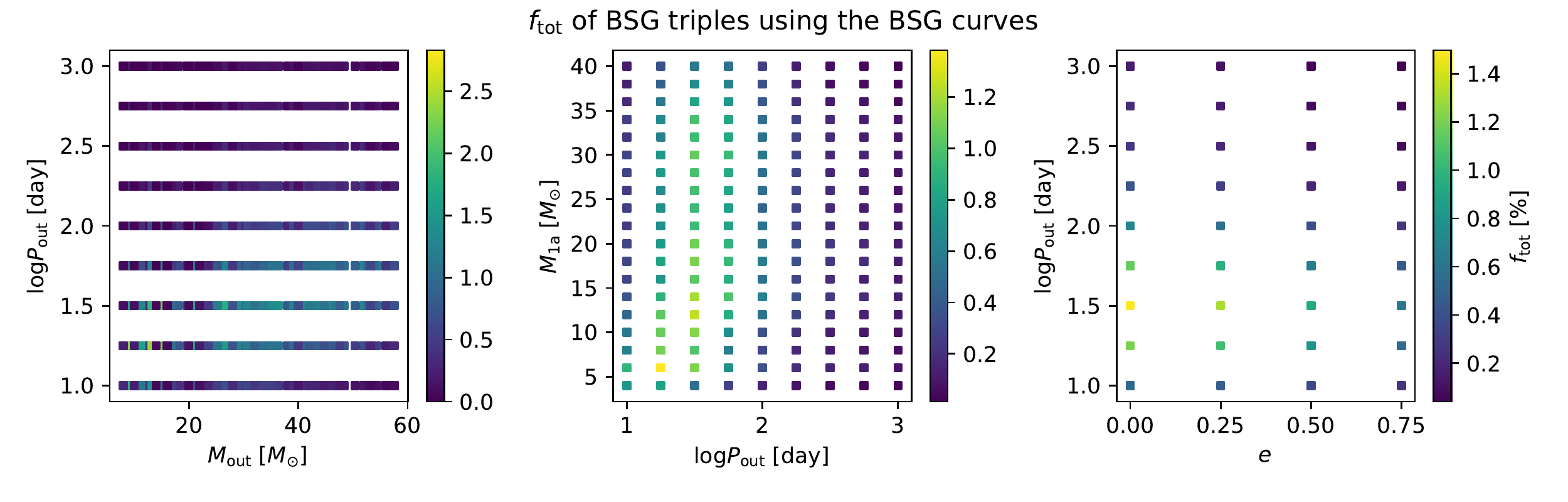}
    \end{subfigure}
    \caption{Same as \Figref{fig_grid_identified_part_SG}, but for more parameters.}
    \label{fig_grid_SGs_full}
\end{figure*}

Figures \ref{fig_grid_OBs_full} and \ref{fig_grid_SGs_full} show the total identification fractions for the triples investigated in \Secref{sec_triple_grid}, but for more parameters such as the eccentricity $e$.

\end{appendices}

\end{document}